\documentclass[onecolumn, prd, aps, tightenlines, preprintnumbers, showpacs, nofootinbib, superscriptaddress, notitlepage]{revtex4-1}

\pdfoutput=1

\usepackage{float}
\usepackage{amsmath}
\usepackage{color}
\usepackage{graphicx}
\usepackage[dvipsnames]{xcolor}
\usepackage{url}
\usepackage{epsfig}
\usepackage[T1]{fontenc}
\usepackage{multirow}
\usepackage{physics} 
\usepackage{booktabs} 
\usepackage{array} 
\usepackage{paralist} 
\usepackage{grffile}
\usepackage{verbatim} 
\usepackage{subfig} 
\usepackage{amsmath,amsthm,amssymb,bm,amsfonts}
\usepackage{slashed}
\usepackage[utf8]{inputenc}
\usepackage{hyperref}
\allowdisplaybreaks[1]

\usepackage{color}
\usepackage[normalem]{ulem}

\def\beg{\begin{equation}}
\def\eeg{\end{equation}}
\def\bea{\begin{eqnarray}}
\def\eea{\end{eqnarray}}

\usepackage{multirow}
\usepackage[title]{appendix}

\def\Tr{{\rm tr}}

\newcommand{\slv}{\raise.15ex\hbox{$/$}\kern-.53em\hbox{$v$}}
\newcommand{\slnbar}{\raise.15ex\hbox{$/$}\kern-.53em\hbox{$\bar{n}$}}
\newcommand{\slF}{\raise.15ex\hbox{$/$}\kern-.53em\hbox{$F$}}
\newcommand{\sllbar}{\raise.15ex\hbox{$/$}\kern-.40em\hbox{$\bar{l}$}}
\newcommand{\slh}{\raise.15ex\hbox{$/$}\kern-.40em\hbox{$h$}}
\newcommand{\slP}{\raise.15ex\hbox{$/$}\kern-.53em\hbox{$P$}}
\newcommand{\slR}{\raise.15ex\hbox{$/$}\kern-.53em\hbox{$R$}}
\newcommand{\slz}{\raise.15ex\hbox{$/$}\kern-.53em\hbox{$Z$}}
\newcommand{\slzbar}{\raise.15ex\hbox{$/$}\kern-.53em\hbox{$\bar{Z}$}}
\newcommand{\slQ}{\raise.15ex\hbox{$/$}\kern-.53em\hbox{$Q$}}
\newcommand{\slK}{\raise.15ex\hbox{$/$}\kern-.53em\hbox{$K$}}
\newcommand{\slkbar}{\raise.15ex\hbox{$/$}\kern-.53em\hbox{$\bar{k}$}}
\newcommand{\slkone}{\raise.15ex\hbox{$/$}\kern-.53em\hbox{$k_1$}}
\newcommand{\slpone}{\raise.15ex\hbox{$/$}\kern-.53em\hbox{$p_1$}}
\newcommand{\slpbarone}{\raise.15ex\hbox{$/$}\kern-.53em\hbox{$\bar{p}_1$}}
\newcommand{\slptwo}{\raise.15ex\hbox{$/$}\kern-.53em\hbox{$p_2$}}
\newcommand{\slpbartwo}{\raise.15ex\hbox{$/$}\kern-.53em\hbox{$\bar{p}_2$}}
\newcommand{\slqone}{\raise.15ex\hbox{$/$}\kern-.53em\hbox{$q_1$}}
\newcommand{\slD}{\raise.15ex\hbox{$/$}\kern-.53em\hbox{$\!D$}}
\newcommand{\slC}{\raise.15ex\hbox{$/$}\kern-.53em\hbox{$C$}}
\newcommand{\slA}{\raise.15ex\hbox{$/$}\kern-.73em\hbox{$A$}}
\newcommand{\slSigma}{\raise.15ex\hbox{$/$}\kern-.53em\hbox{$\Sigma$}}
\newcommand{\slpartial}{\raise.15ex\hbox{$/$}\kern-.53em\hbox{$\partial$}}
\newcommand{\slcalP}{\raise.15ex\hbox{$/$}\kern-.63em\hbox{$\cal P$}}
\newcommand{\sleps}{\raise.15ex\hbox{$/$}\kern-.53em\hbox{$\epsilon$}}
\newcommand{\slepsbar}{\raise.15ex\hbox{$/$}\kern-.53em\hbox{$\overline{\epsilon}$}}
\newcommand{\slepsstar}{\raise.15ex\hbox{$/$}\kern-.53em\hbox{$\epsilon$}^\star}
\newcommand{\slS}{\raise.15ex\hbox{$/$}\kern-.73em\hbox{$S$}}

\newcommand{\bb}{\mathbf}

\newcommand{\be}{\boldsymbol{\epsilon}}
\newcommand{\bk}{\mathbf{k}}
\newcommand{\bp}{\mathbf{p}}
\newcommand{\bq}{\mathbf{q}}
\newcommand{\bx}{\mathbf{x}}

\newcommand{\barq}{\bar{q}}
\newcommand{\de}{\cdot\boldsymbol{\epsilon}}
\newcommand{\des}{\cdot \boldsymbol{\epsilon}^*}

\newcommand{\dtwo}[1]{\frac{\dd^2 #1}{(2\pi)^2}}

\newcommand{\p}{\prime}

\newcommand{\zho}{z_{h_1}}
\newcommand{\zht}{z_{h_2}}

\newcommand{\eq}[1]{\begin{align} #1 \end{align}}

\newcommand{\pa}[1]{\left( #1 \right)}
\newcommand{\br}[1]{\left[ #1 \right]}

\begin{document}
\title{Dihadron production in DIS at small $x$ at next to leading order: transverse photons}

\author{Filip Bergabo}
\email{fbergabo@gradcenter.cuny.edu}
\affiliation{Department of Natural Sciences, Baruch College, CUNY, 17 Lexington Avenue, New York, NY 10010, USA}
\affiliation{City University of New York Graduate Center, 365 Fifth Avenue, New York, NY 10016, USA}

\author{Jamal Jalilian-Marian}
\email{jamal.jalilian-marian@baruch.cuny.edu}
\affiliation{Department of Natural Sciences, Baruch College, CUNY, 17 Lexington Avenue, New York, NY 10010, USA}
\affiliation{City University of New York Graduate Center, 365 Fifth Avenue, New York, NY 10016, USA}

\begin{abstract}
We calculate the next to leading order corrections to dihadron production in Deep Inelastic Scattering (DIS) at small $x$ using the Color Glass Condensate formalism for the case when the virtual photon is transverse polarized. Similar to the case of longitudinal photon exchange all UV and soft singularities cancel while the collinear divergences are absorbed into quark and antiquark-hadron fragmentation functions. Rapidity divergences lead to JIMWLK evolution of dipoles and quadrupoles which describe multiple-scatterings of the quark antiquark dipole on the target proton/nucleus and contain all the QCD dynamics of the target leading to a finite final result for the dihadron production cross section.
\end{abstract}

\maketitle



\section{Introduction}\label{sec:intro}
A hadron or nucleus wave function at high energy (equivalently, small $x$) contains a 
large number of predominantly gluons leading to the phenomenon of gluon 
saturation~\cite{Iancu:2003xm,Jalilian-Marian:2005ccm,Weigert:2005us,Gelis:2010nm,Morreale:2021pnn}. 
Inclusive and diffractive two-particle production and angular correlations in high energy 
hadronic/nuclear collisions is a sensitive probe of gluon saturation in a proton or nucleus at small 
$x$~\cite{Kovner:2001vi,JalilianMarian:2004da,Marquet:2007vb,Albacete:2010pg,Stasto:2011ru,Lappi:2012nh,Stasto:2018rci,Albacete:2018ruq,Boussarie:2021lkb,Fujii:2020bkl,Kotko:2015ura,vanHameren:2016ftb,Altinoluk:2021ygv,Hatta:2020bgy,Jia:2019qbl,Jalilian-Marian:2012wwi,Jalilian-Marian:2011tvq,Jalilian-Marian:2005tod,Jalilian-Marian:2004cdc,Dumitru:2011zz,Dumitru:2010ak,Kang:2011bp,Kolbe:2020tlq,Jalilian-Marian:2005qbq,Mantysaari:2019hkq,Boussarie:2021ybe,Kotko:2017oxg,Salazar:2019ncp,Mantysaari:2019csc,Altinoluk:2015dpi,Dumitru:2015gaa,Iancu:2021rup,Hatta:2016dxp,Hatta:2020bgy,Boussarie:2016ogo,Boussarie:2014lxa}.
The disappearance of the away side peak in proton (deuteron)-nucleus collisions in the 
forward rapidity region at RHIC~\cite{Braidot:2010ig,Adare:2011sc} as predicted by gluon saturation 
models~\cite{Marquet:2007vb} provides the strongest hint for 
the presence of gluon saturation in the wave function of the target nucleus at small $x$. 
Nevertheless due to complications arising from further interactions and radiation from 
both initial and final states an unambiguous interpretation of the RHIC results remains 
illusive. DIS offers the cleanest environment in which the dynamics of gluon saturation can be investigated theoretically as the virtual photon probing the inner structure of the target does not interact strongly. The proposed Electron-Ion Collider (EIC) will allow precision studies of the observables~\cite{Aschenauer:2016our,Accardi:2012qut} in which gluon saturation is expected to play a dominant role and as such establish the presence of saturation and clarify the kinematics in which it is the main QCD effect. Due to this fact it is imperative that the existing predictions for saturation effects are made more precise by calculating higher orders in $\alpha_s$ corrections. 

Next to leading order calculations for many processes using the Color Glass Condensate effective theory of QCD at small $x$ have recently become available~\cite{Chirilli:2011km,Chirilli:2012jd,Ayala:2016lhd,Ayala:2017rmh,Caucal:2021ent,Caucal:2022ulg,Bergabo:2021woe,Bergabo:2022tcu,Taels:2022tza,Iancu:2020mos,Bergabo:2022zhe,Benic:2016uku}. In a recent paper~\cite{Bergabo:2022tcu} we calculated the one-loop corrections to inclusive dihadron production in DIS at small $x$ for the case when the exchanged virtual photon is longitudinal. In this work we extend our studies of this process and calculate dihadron production in DIS with transverse photon echange. As the calculational methods are identical to our earlier work we will skip a lot of the details of the calculation and refer the reader to~\cite{Bergabo:2022tcu}. As before there are several divergences that appear at the next to leading order. All divergences either cancel or can be absorbed into evolution of physical quantities. Our final results are then completely finite and can be used to calculate inclusive dihadron production and angular correlations in DIS at small $x$. 

In the small $x$ limit of DIS the virtual photon (transverse or longitudinal) splits into a quark antiquark pair (a dipole), which then multiply scatters from the target hadron or nucleus. To leading order (LO)  accuracy the double inclusive production cross section can be written as  

\bea
\frac{\dd \sigma^{\gamma^*A \to q\bar{q} X}}{\dd^2 \bb{p}\, \dd^2 \bb{q} \, \dd y_1 \, \dd y_2} &=& \frac{ e^2 Q^2(z_1z_2)^2 N_c}{(2\pi)^7} \delta(1-z_1-z_2)\int \dd^8 \bx \left[S_{122^\prime 1^\prime} - S_{12} - S_{1^\prime 2^\prime} + 1\right] \nonumber \\
&& e^{i\bb{p}\cdot(\bb{x}_1^\prime - \bb{x}_1)} e^{i\bb{q}\cdot(\bb{x}_2^\prime - \bb{x}_2)} 
\bigg\{4z_1z_2K_0(|\bb{x}_{12}|Q_1)K_0(|\bb{x}_{1^\prime 2^\prime}|Q_1) + \nonumber \\
&&  (z_1^2 + z_2^2) \,
\frac{ \bb{x}_{12}\cdot \bb{x}_{1^\prime 2^\prime}}{|\bb{x}_{12}| |\bb{x}_{1^\prime 2^\prime}|} \, 
K_1(|\bb{x}_{12}|Q_1)K_1(|\bb{x}_{1^\prime 2^\prime}|Q_1) 
\bigg\} .\label{LOdsig}
\eea
where the first and second terms in the curly bracket above correspond to the contribution of the longitudinal and transverse polarizations of the virtual photon. The production cross section is a convolution of the probability for a photon to split into a quark at transverse position $\bx_1$ and an anti-quark at position $\bx_2$ represented by the Bessel functions, with the probability for this quark antiquark pair to scatter from the target encoded in the dipoles $S_{ij}$ and quadrupoles $S_{ijkl}$. The virtual photon has momentum $l^\mu$ with $l^2 = -Q^2$ and we have set the transverse momentum of the photon to zero without any loss of generality. Furthermore $p^\mu$ ($q^\mu$) is the momentum of the outgoing quark (antiquark) and $z_1$ ($z_2$) is its longitudinal momentum fraction relative to the photon. $\bx_1$ ($\bx_2$) is the transverse coordinate of the quark (antiquark), and primed coordinates are used in the conjugate amplitude. Quark and antiquark rapidities $y_1$ and $y_2$ are related to their momentum fractions $z_1$ and $z_2$ via $\dd y_i = \dd z_i / z_i$. For convenience we also define and use the following shorthand notations:

\begin{align}
Q_i = Q\sqrt{z_i(1-z_i)}, \,\,\,\,\,\, \bx_{ij} = \bx_i - \bx_j,\,\,\,\,\,\, \dd^8 \bx = \dd^2 \bx_1 \, \dd^2 \bx_2\, \dd^2 \bx_{1^\p} \, \dd^2 \bx_{2^\p}.
\end{align}
Dipoles $S_{ij}$ and quadrupoles $S_{ijkl}$ are normalized correlation functions of two and four 
Wilson lines defined as 

\begin{align}
S_{ij} = \frac{1}{N_c} \Tr\left\langle V_i V_j^\dag \right\rangle, \,\,\,\,\,\,\,\,\, S_{ijkl} = \frac{1}{N_c}\Tr\left\langle V_i V_j^\dag V_k V_l^\dag\right\rangle, \label{dipquad}
\end{align}
which contain the full dynamics of gluon saturation. Here index $i$ refers to the transverse coordinate $\bb{x}_i$ and we use the following notation for Wilson lines 
\begin{align}
V_i &= \hat{P}\exp\left( ig \int \dd x^+ A^-(x^+,\bb{x}_i)\right),
\end{align}
which resum the multiple scatterings of the quark and antiquark from the target hadron or nucleus.

S

\section{One-loop corrections}

\begin{figure}[H]
\centering
\includegraphics[width=70mm]{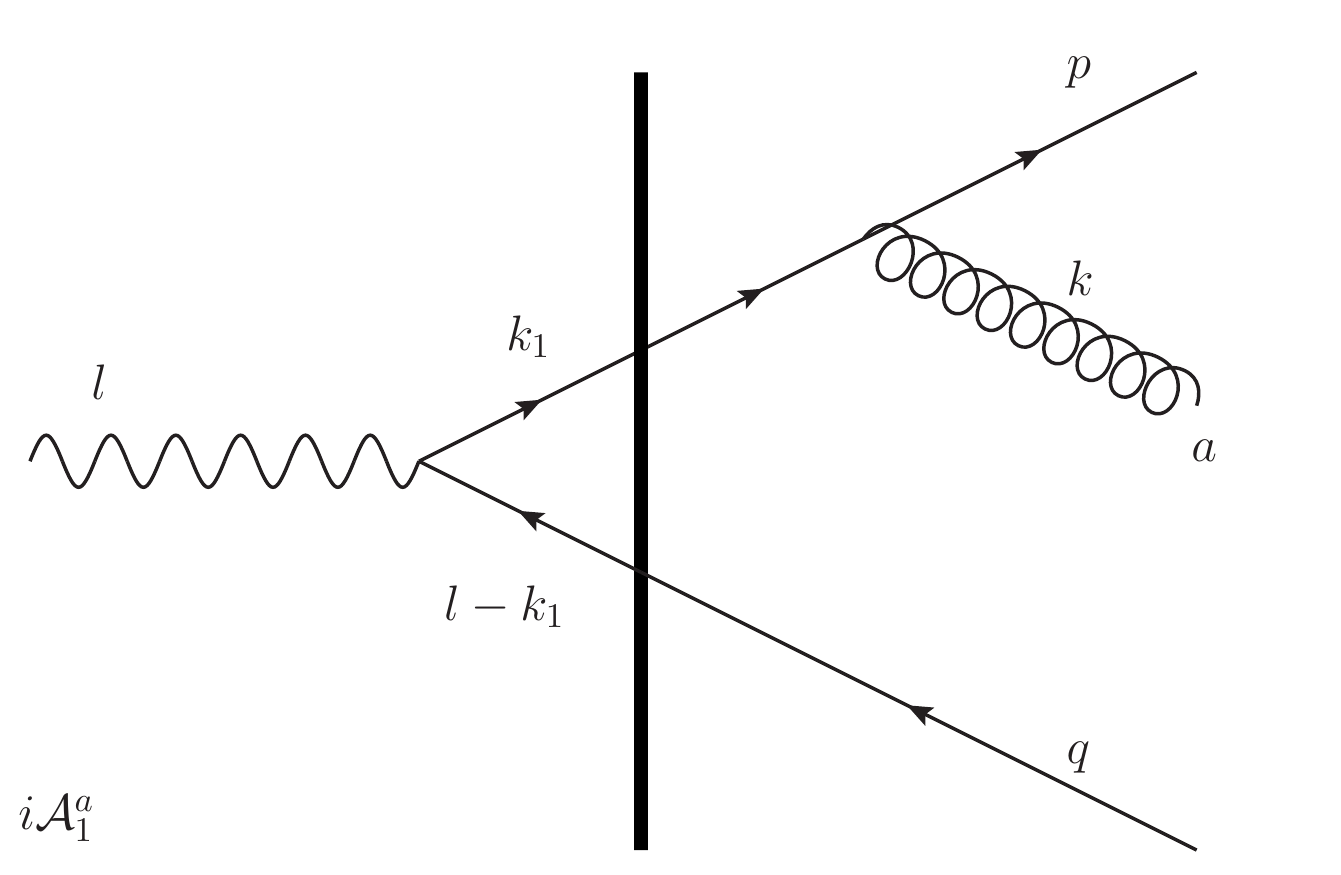}\includegraphics[width=70mm]{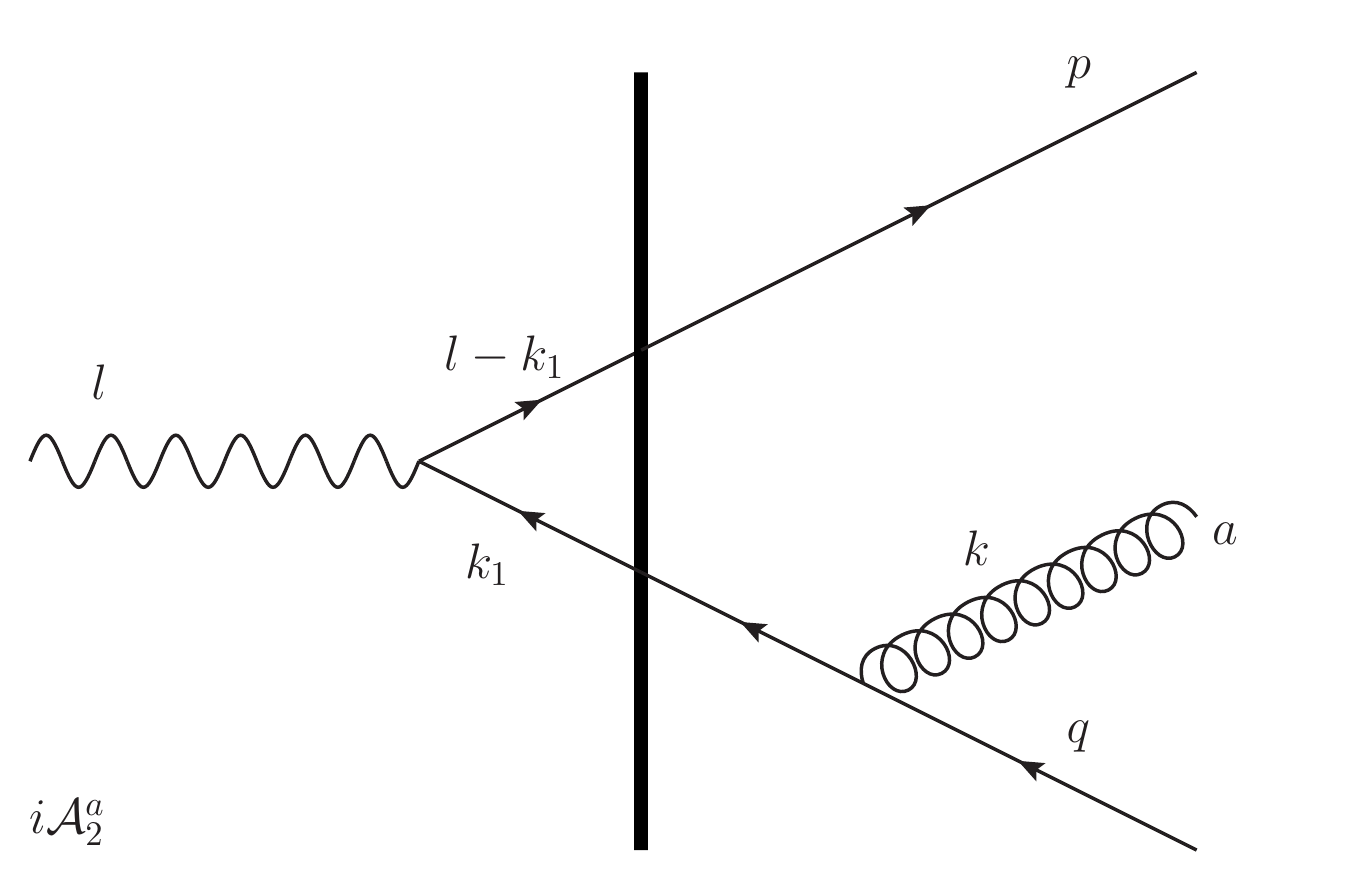}\\ \includegraphics[width=70mm]{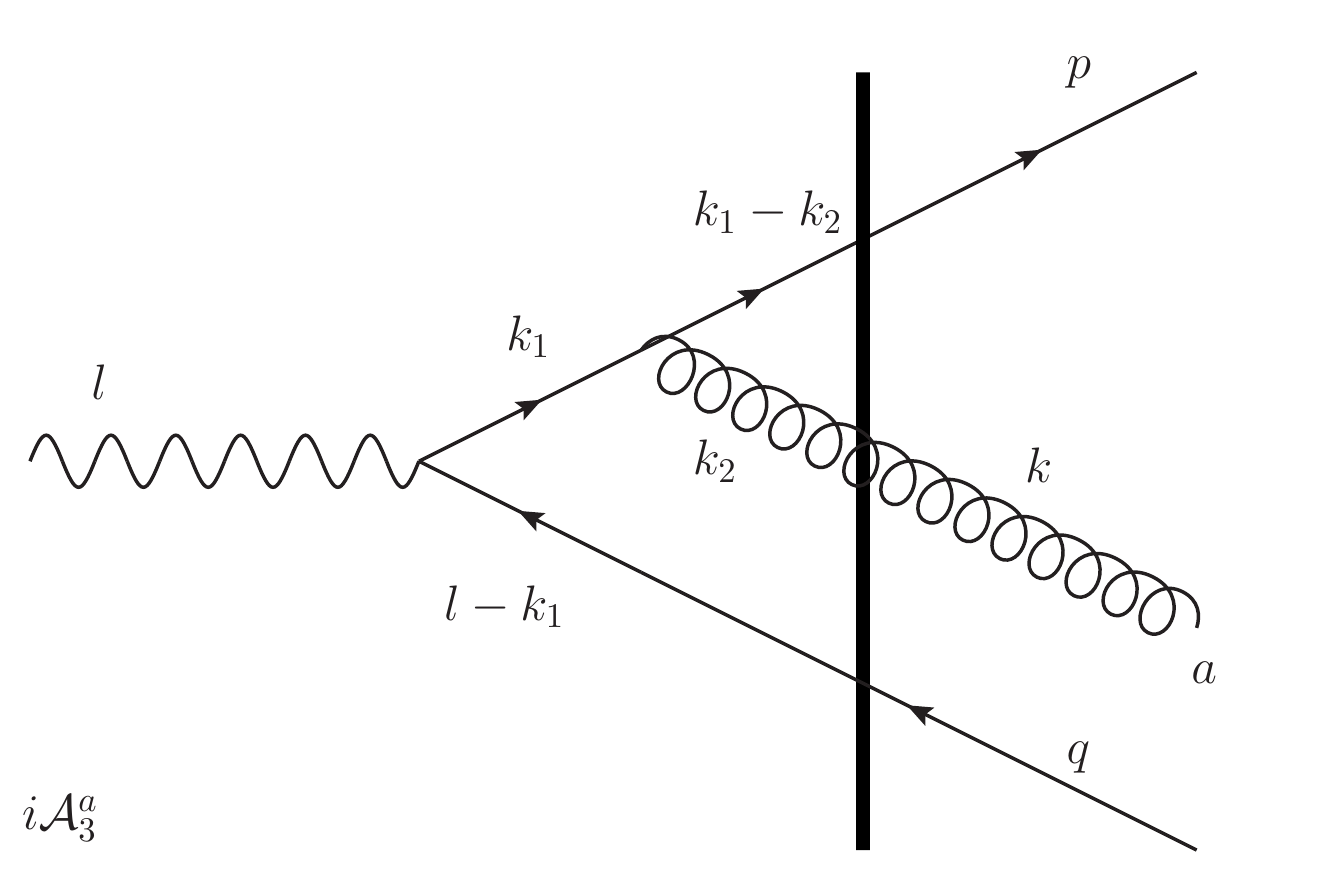}\includegraphics[width=70mm]{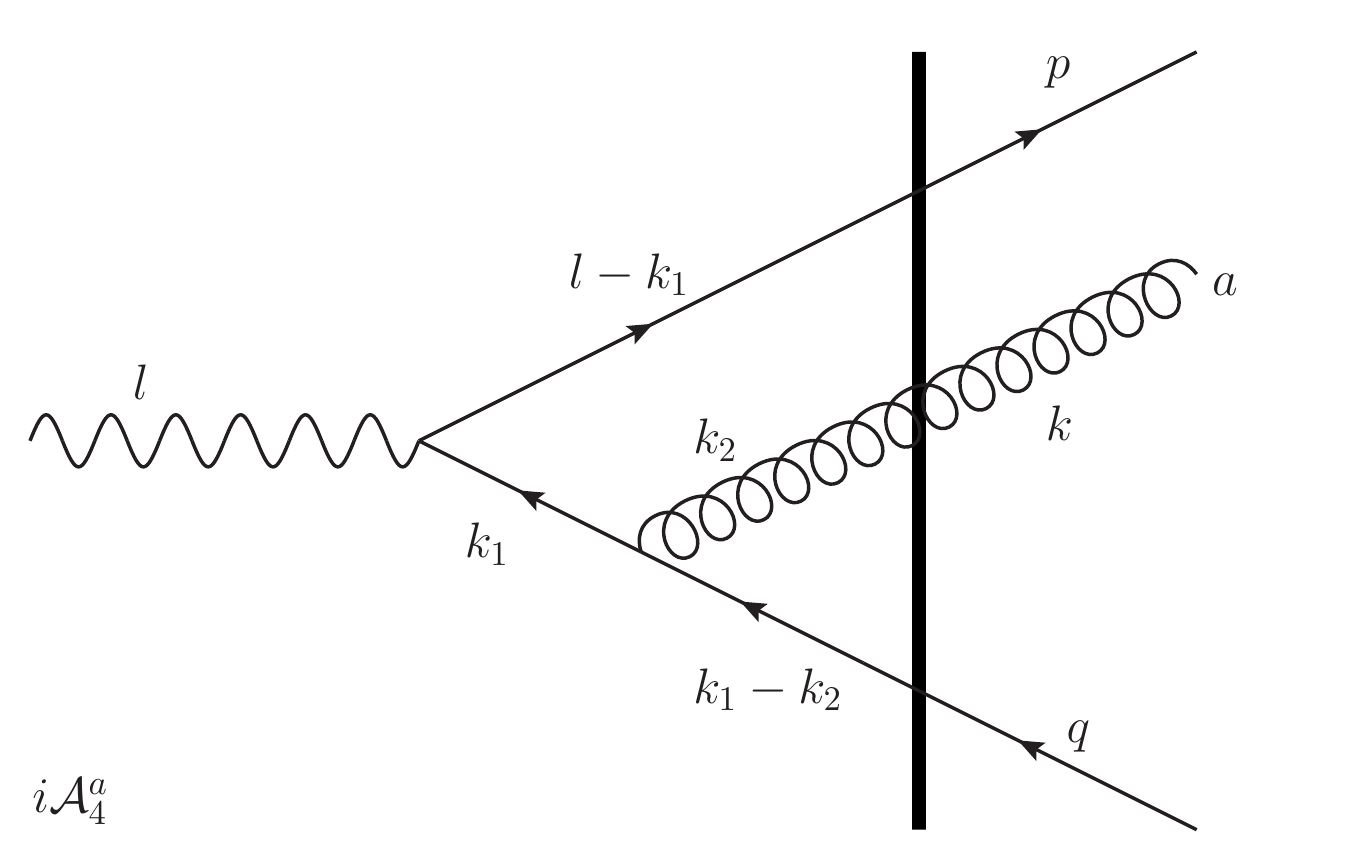}
\caption{The real corrections $i\mathcal{A}_1^a, ..., i\mathcal{A}_4^a$. The arrows on Fermion lines indicate Fermion number flow, all momenta flow to the right. The thick solid line indicates interaction with the target.}\label{fig:realdiags}
\end{figure}

\begin{figure}[H]
\centering
\includegraphics[width=60mm]{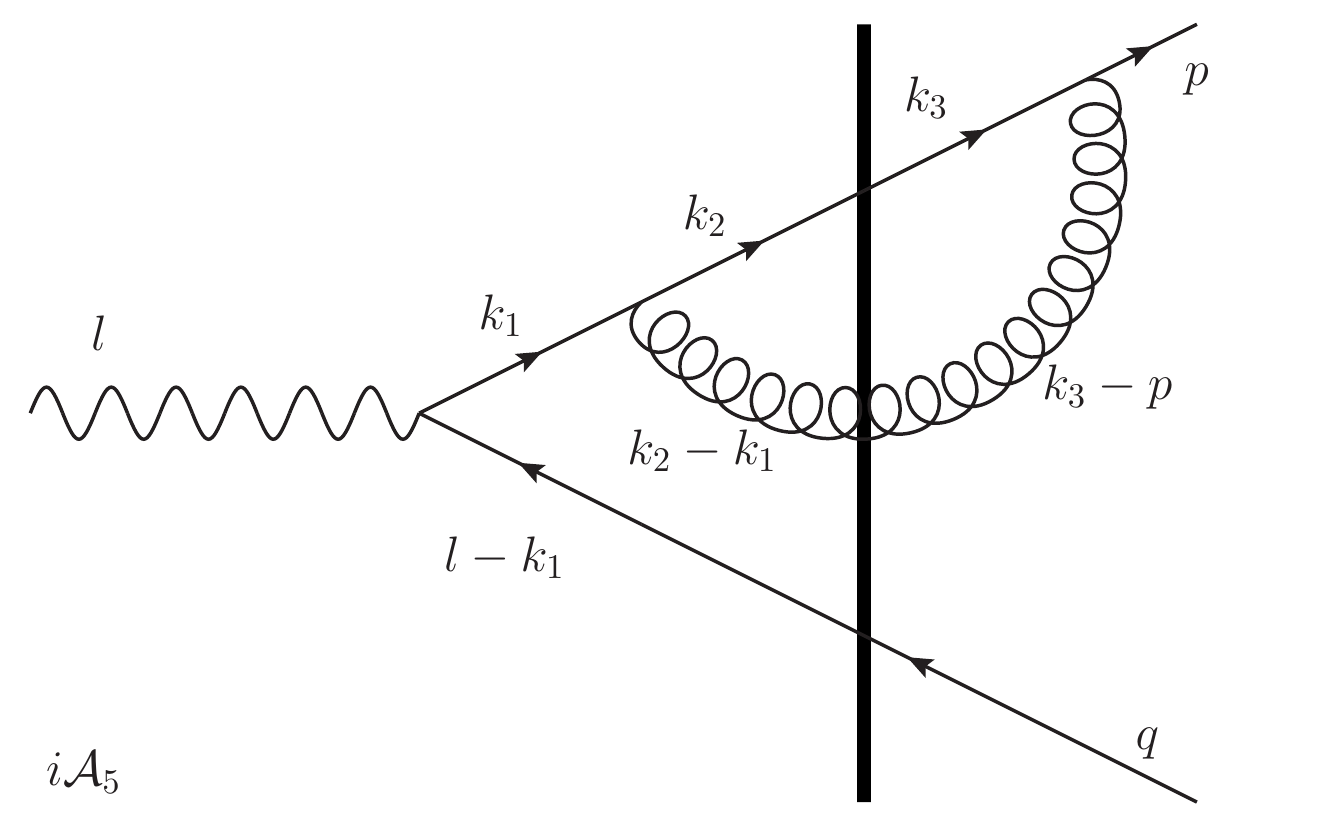}\includegraphics[width=60mm]{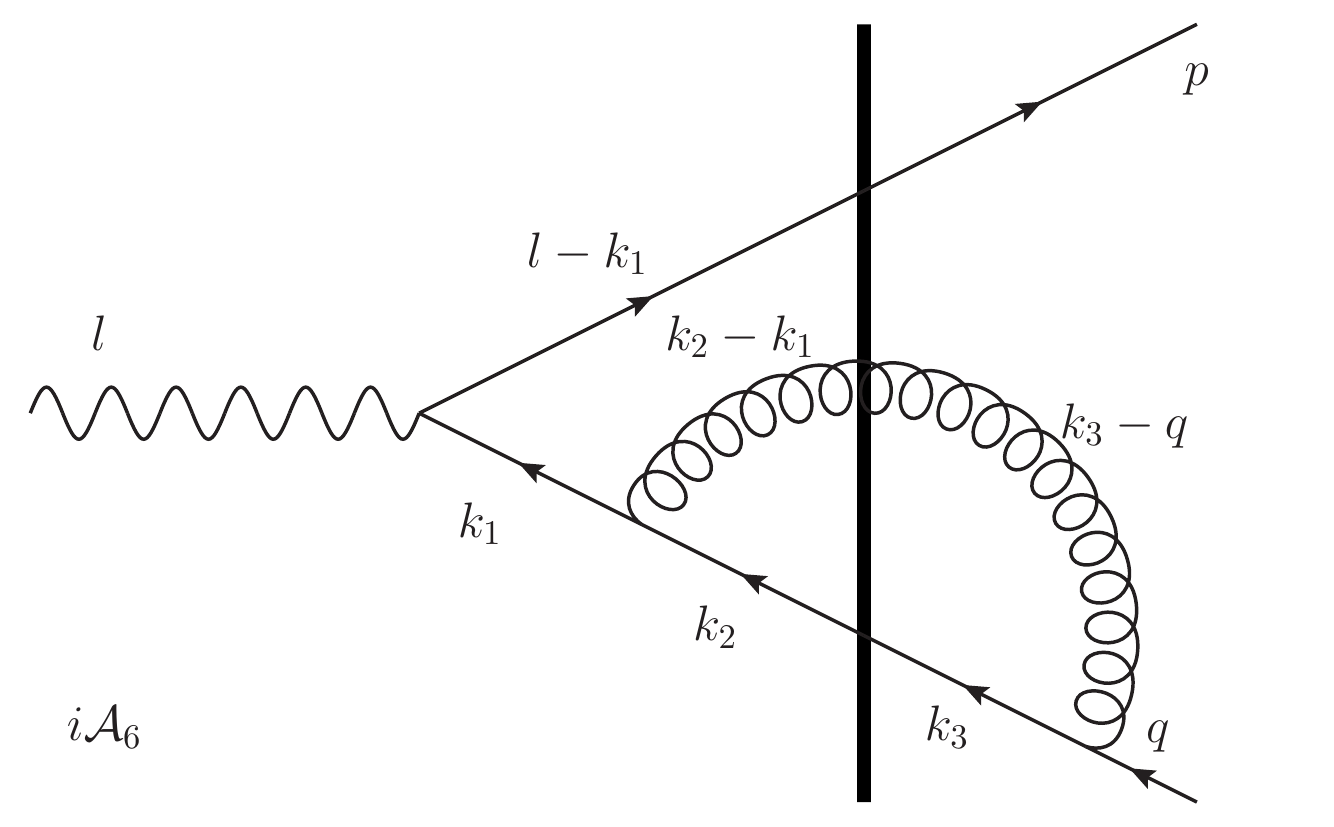}\\ \includegraphics[width=60mm]{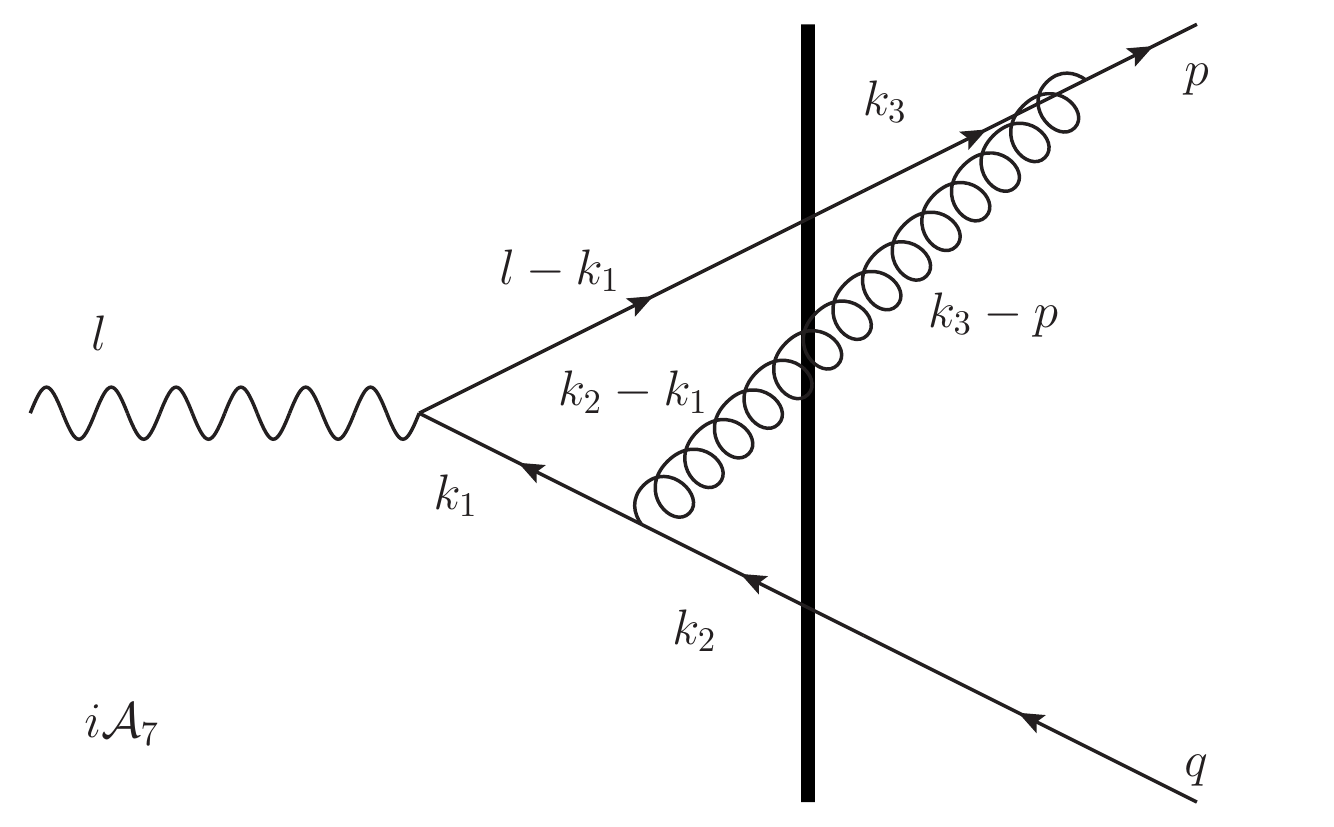}\includegraphics[width=60mm]{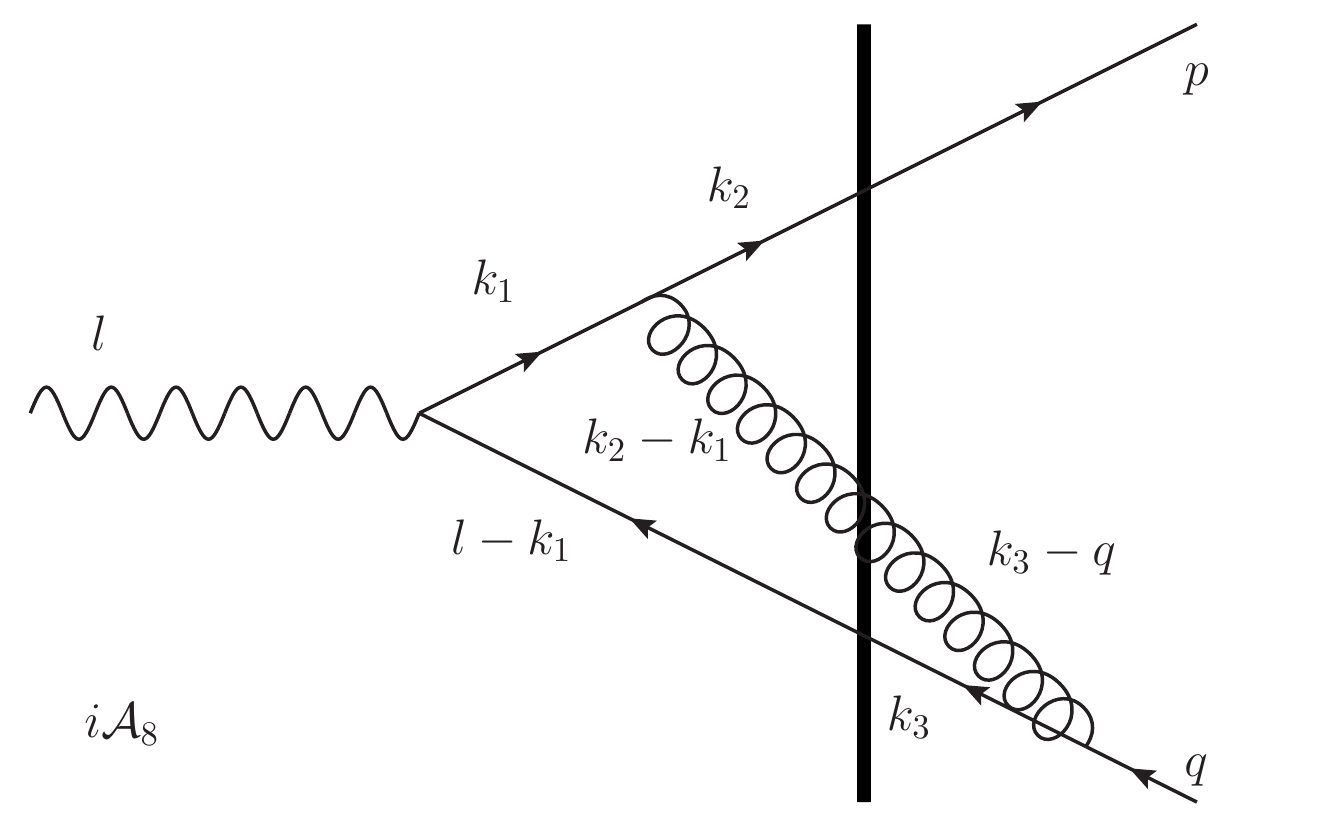}\\ \includegraphics[width=60mm]{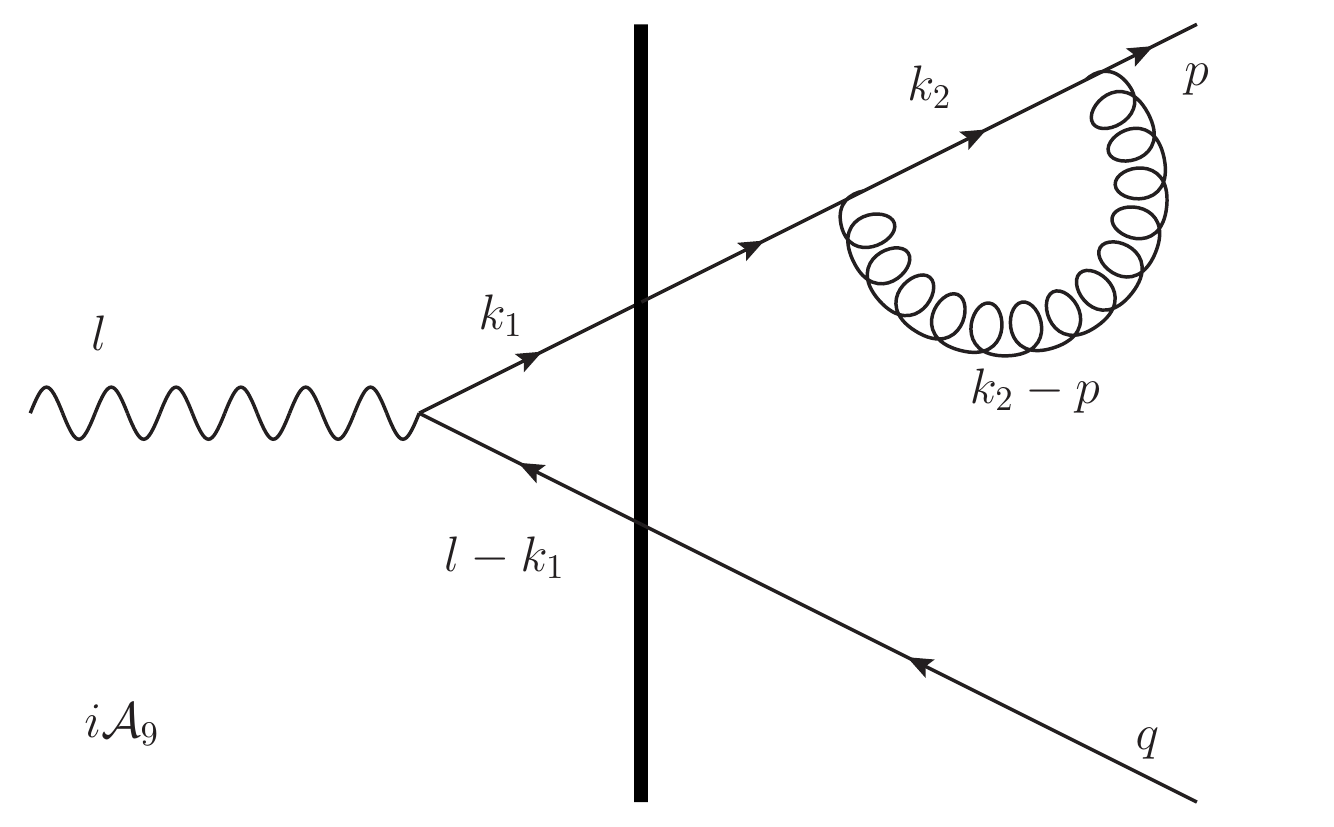}\includegraphics[width=60mm]{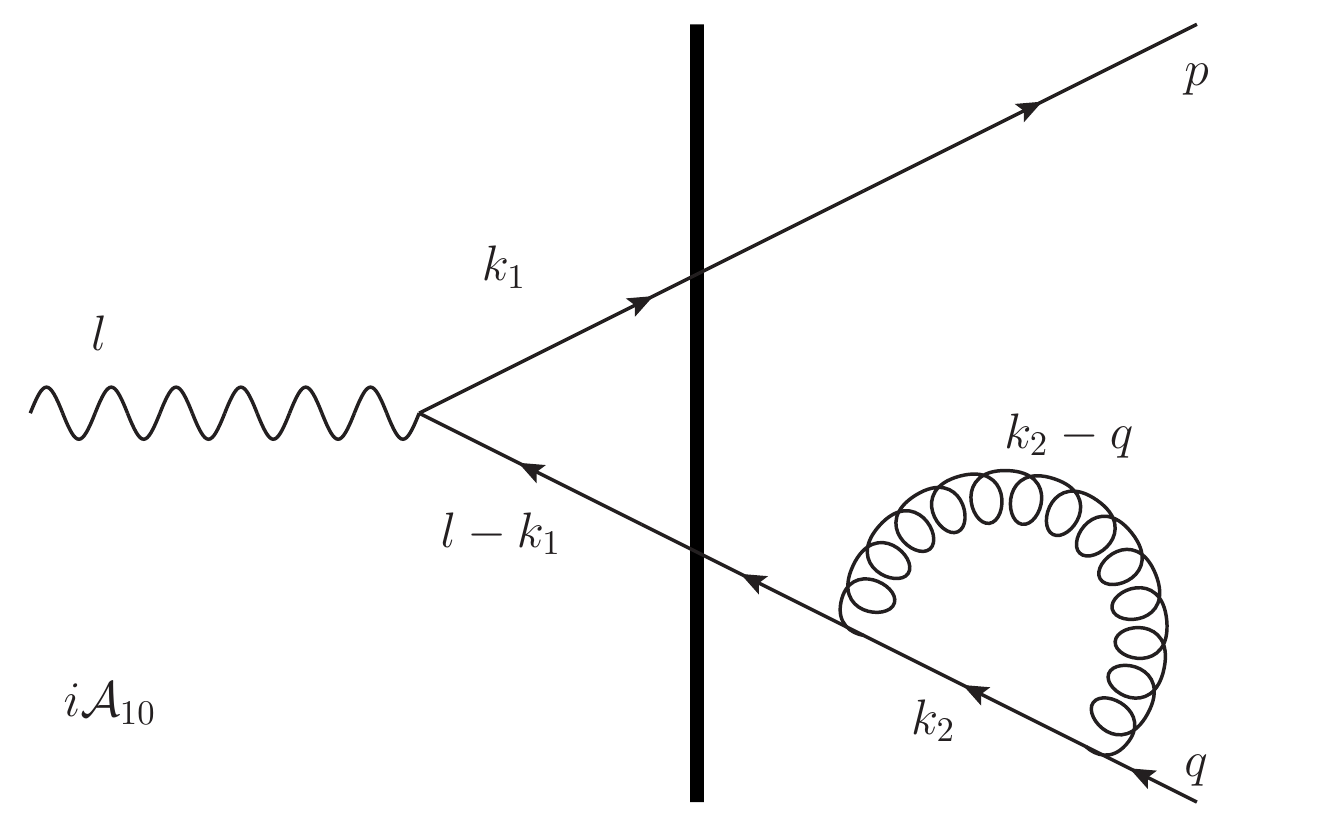}\\ \includegraphics[width=60mm]{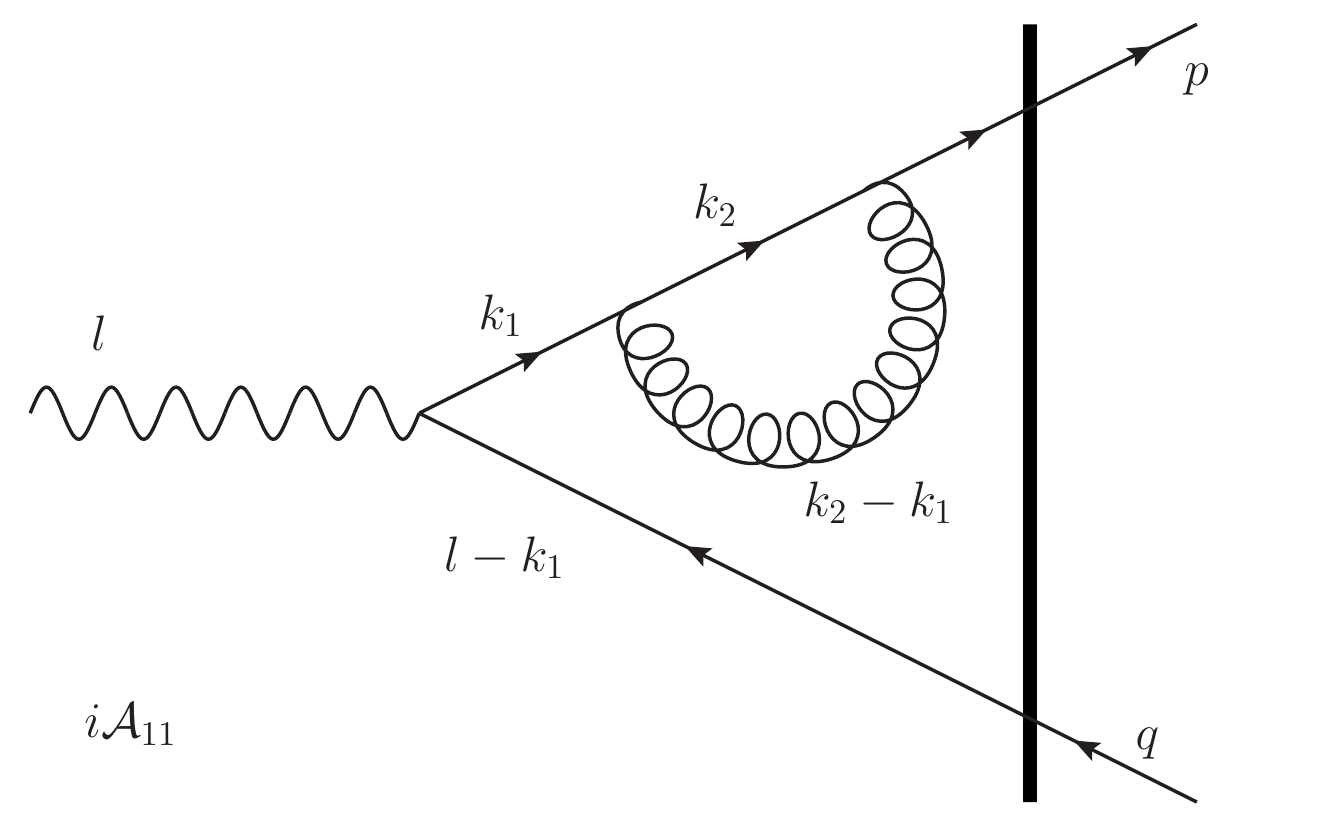}\includegraphics[width=60mm]{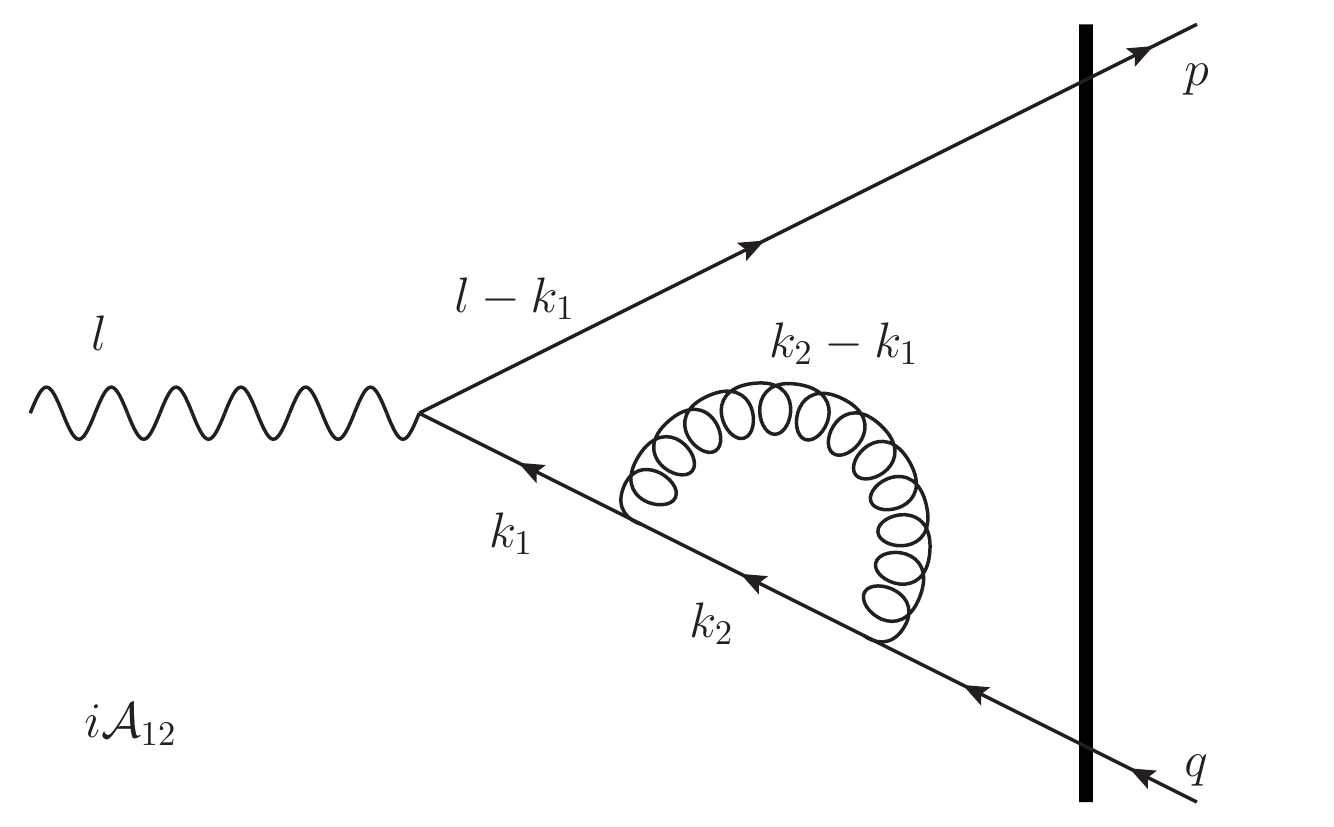}\\ \includegraphics[width=60mm]{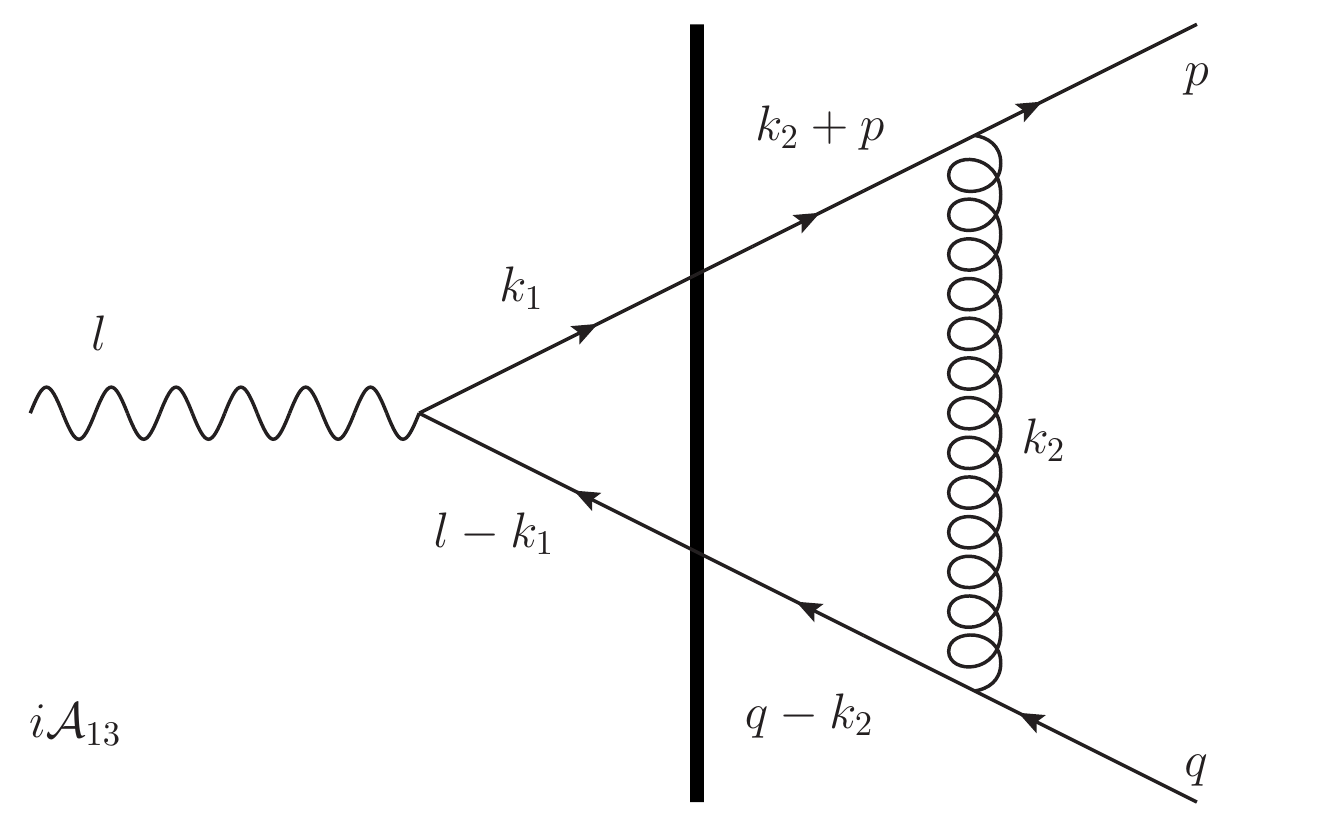}\includegraphics[width=60mm]{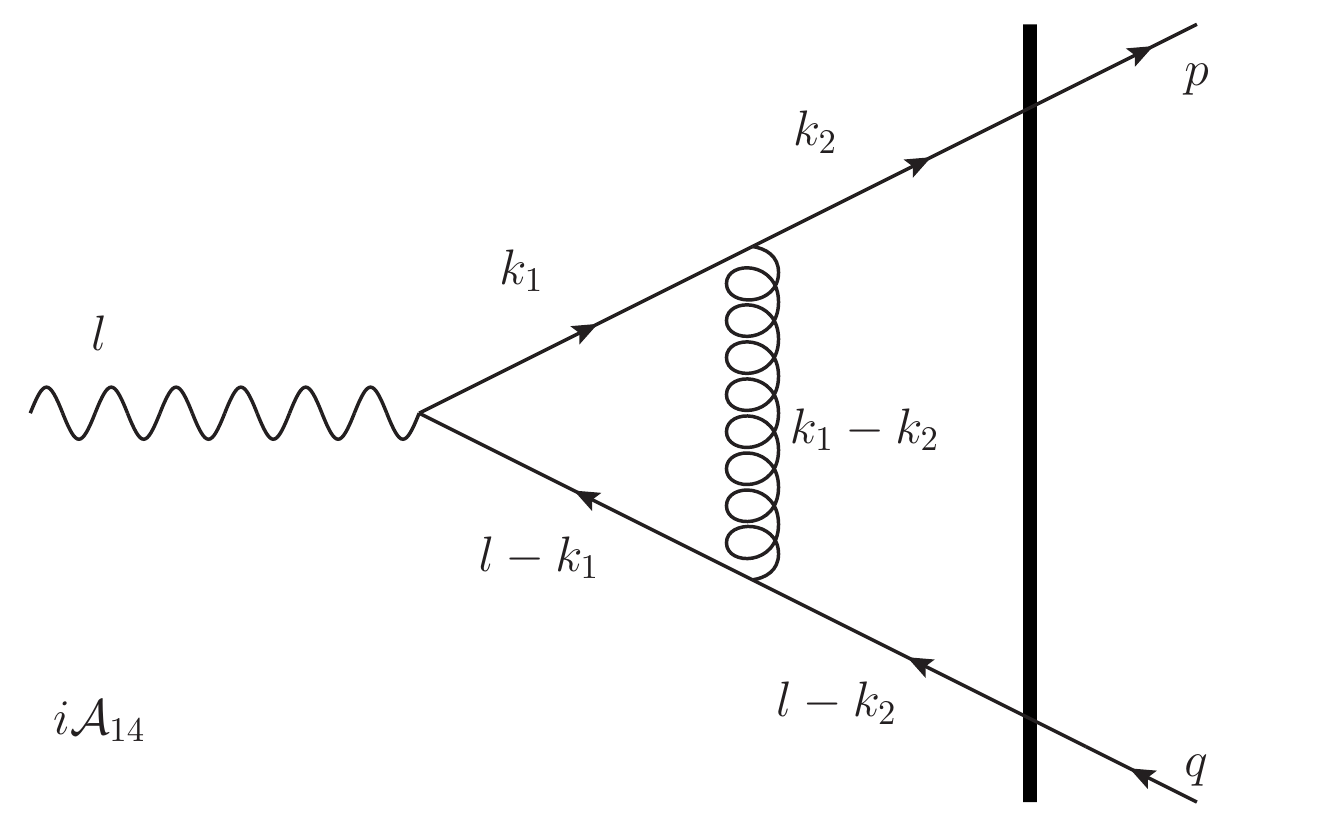}
\caption{The ten virtual NLO diagrams $i\mathcal{A}_5, ..., i\mathcal{A}_{14}$. The arrows on fermion lines indicate fermion number flow, all momenta flow to the right, \textit{except} for gluon momenta. The thick solid line indicates interaction with the target.}\label{virtualdiags}
\end{figure}

In~\cite{Ayala:2016lhd,Ayala:2017rmh, Bergabo:2021woe, Bergabo:2022tcu, Bergabo:2022zhe} 
spinor helicity formalism was used to calculate the contribution of real diagrams to next 
to leading order corrections to the leading order results. The real corrections are shown 
in Fig. (\ref{fig:realdiags}) and involve radiation of a gluon either by the quark or 
antiquark before they scatter from the target in which case the gluon also scatters 
from the target~\cite{Dumitru:2002qt,Dumitru:2001jn,Ayala:1995hx}, or after they scatter 
from the target in which 
case the radiated gluon does not scatter from the target,
The virtual corrections are shown in figure \ref{virtualdiags} and involve radiation of a gluon by either quark or antiquark which is then absorbed by the quark or antiquark line still in the amplitude~\cite{Bergabo:2022tcu, Bergabo:2022zhe}, 

We refer the reader to~\cite{Bergabo:2022tcu} for the explicit expressions for the amplitudes 
for real and virtual corrections given by eqs. ($5$) and ($8-14$) respectively, and 
eqs. ($6$) and ($14-19$) for the Dirac numerators. In this study, we focus on the contribution from transversely polarized photons and we compute the numerators using the spinor helicity formalism. The needed real numerators are in table \ref{numtable} and the virtual numerators are in Eq. \ref{N5} - \ref{N14}.

\subsection{Dirac numerators for real diagrams}

\begin{table}[H]
\centering
\begin{tabular}{| c
 | c | l |}
\hline
Numerator & $\lambda_\gamma ; \lambda_q ,\lambda_g$ & $N_i^{\lambda_\gamma ; \lambda_q, \lambda_g}$ \\[2pt]
\hline
\multirow{6}{1em}{\vspace{-1ex}$N_1$}
& $+;+,+$ & $ -(z_1)^{3/2}\sqrt{z_2}(1-z_2)\frac{[(z_1\bk-z_3\bp)\de]}{(z_1\bk-z_3\bp)^2}\, (\bk_1\de)$ \\ [5pt] 
& $+;+,-$ & $- \sqrt{z_1z_2}(1-z_2)^2\frac{[(z_1\bk-z_3\bp)\des]}{(z_1\bk-z_3\bp)^2}\,(\bk_1\de)$ \\[5pt] 
& $+;-,+$ & $ (z_2)^{3/2}\sqrt{z_1}(1-z_2)\frac{[(z_1\bk-z_3\bp)\de]}{(z_1\bk-z_3\bp)^2}\,(\bk_1\de)$ \\[5pt] 
& $+,-,-$ &   $(z_1z_2)^{3/2}\frac{[(z_1\bk-z_3\bp)\des]}{(z_1\bk-z_3\bp)^2}\, (\bk_1\de)$\\[5pt] 
\hline
\multirow{6}{1em}{\vspace{-1ex}$N_2$} 
& $+;+,+$ & $ -  (z_1)^{3/2}\sqrt{z_2} (1-z_1) \frac{[(z_2\bk-z_3\bq)\de]}{(z_2\bk-z_3\bq)^2}\,(\bk_1\de)$ \\ [5pt] 
& $+;+,-$ & $ - (z_1z_2)^{3/2}  \frac{[(z_2\bk-z_3\bq)\des]}{(z_2\bk-z_3\bq)^2}\, (\bk_1\de)$ \\[5pt] 
& $+;-,+$ & $ (z_2)^{3/2}\sqrt{z_1}(1-z_1)  \frac{[(z_2\bk-z_3\bq)\de]}{(z_2\bk-z_3\bq)^2}\, (\bk_1\de)$ \\[5pt] 
& $+,-,-$ & $\sqrt{z_1z_2}(1-z_1)^2   \frac{[(z_2\bk-z_3\bq)\des]}{(z_2\bk-z_3\bq)^2} \, (\bk_1\de)$ \\[5pt] 
\hline
\multirow{6}{1em}{\vspace{-1ex}$N_3$}
& $+;+,+$ & $ (z_1)^{3/2}\sqrt{z_2}(1-z_2)\left( \frac{ \mathbf{k}_2 \cdot \boldsymbol{\epsilon}}{z_3} - \frac{ \mathbf{k}_1 \cdot \boldsymbol{\epsilon}}{1-z_2}\right) \mathbf{k}_1 \cdot \boldsymbol{\epsilon}$ \\ [5pt] 
& $+;+,-$ & $\sqrt{z_1z_2}(1-z_2)^2  \left( \frac{ \mathbf{k}_2 \cdot \boldsymbol{\epsilon}^*}{z_3} - \frac{ \mathbf{k}_1 \cdot \boldsymbol{\epsilon}^*}{1-z_2}\right)\mathbf{k}_1 \cdot \boldsymbol{\epsilon}$ \\[5pt] 
& $+;-,+$ & $-(z_2)^{3/2}\sqrt{z_1}(1-z_2) \left( \frac{ \mathbf{k}_2 \cdot \boldsymbol{\epsilon}}{z_3} - \frac{ \mathbf{k}_1 \cdot \boldsymbol{\epsilon}}{1-z_2}\right) \mathbf{k}_1 \cdot \boldsymbol{\epsilon}$ \\[5pt] 
& $+,-,-$ & $-(z_1z_2)^{3/2}\left[ \left( \frac{ \mathbf{k}_2 \cdot \boldsymbol{\epsilon}^*}{z_3} - \frac{ \mathbf{k}_1 \cdot \boldsymbol{\epsilon}^*}{(1-z_2)}\right) \mathbf{k}_1 \cdot \boldsymbol{\epsilon} +\frac{\bk_1^2 + z_2(1-z_2)Q^2}{2z_2(1-z_2)}\right]$ \\[5pt] 
\hline
\multirow{6}{1em}{\vspace{-1ex}$N_4$}
& $+;+,+$ & $ (z_1)^{3/2}\sqrt{z_2}(1-z_1)\left( \frac{ \mathbf{k}_2 \cdot \boldsymbol{\epsilon}}{z_3} - \frac{ \mathbf{k}_1 \cdot \boldsymbol{\epsilon}}{1-z_1}\right)\mathbf{k}_1 \cdot \boldsymbol{\epsilon}$ \\ [5pt] 
& $+;+,-$ & $(z_1z_2)^{3/2} \left[ \left(\frac{ \mathbf{k}_2 \cdot \boldsymbol{\epsilon}^*}{z_3} - \frac{  \mathbf{k}_1 \cdot \boldsymbol{\epsilon}^*}{(1-z_1)}\right) \mathbf{k}_1 \cdot \boldsymbol{\epsilon} +\frac{\bk_1^2 + z_1(1-z_1)Q^2}{2z_1(1-z_1)}\right]$ \\[5pt] 
& $+;-,+$ & $-(z_2)^{3/2}\sqrt{z_1}(1-z_1)\left(\frac{ \mathbf{k}_2\cdot \boldsymbol{\epsilon}}{z_3}- \frac{ \mathbf{k}_1 \cdot \boldsymbol{\epsilon}}{1-z_1}\right) \mathbf{k}_1 \cdot \boldsymbol{\epsilon}$ \\[5pt] 
& $+,-,-$ &  $ - \sqrt{z_1z_2}(1-z_1)^2 \left( \frac{ \mathbf{k}_2 \cdot \boldsymbol{\epsilon}^*}{z_3} - \frac{ \mathbf{k}_1 \cdot \boldsymbol{\epsilon}^*}{1-z_1}\right)\mathbf{k}_1 \cdot \boldsymbol{\epsilon}$\\[5pt] 
\hline
\end{tabular}
\caption{The minimal set of transverse photon numerators $N_1$ to $N_4$ in momentum fraction notation. Complex conjugation results in the numerator with all helicities flipped (while leaving longitudinal helicities unchanged), and any numerator where $\lambda_q = \lambda_{\bar{q}}$ is zero.}
\label{numtable}
\end{table}

\begin{align}
N_5^{+;+} =& \frac{2^5 (l^+)^2   z_1^{3/2}\sqrt{z_2}}{(z_1-z_3)^2} \Bigg\{ z_1^2\left[\left(\bk_3-\frac{z_3}{z_1}\bp\right)\de\right]\left[\left(\bk_2-\frac{z_3}{z_1}\bk_1\right)\de^*\right]+ z_3^2\left[\left(\bk_3-\frac{z_3}{z_1}\bp\right)\de^*\right]\left[\left(\bk_2-\frac{z_3}{z_1}\bk_1\right)\de\right]\Bigg\}\bk_1\de, \nonumber \\
N_5^{+;-} =& \frac{-2^5 (l^+)^2   z_2^{3/2}\sqrt{z_1}}{(z_1-z_3)^2} \Bigg\{ z_1^2\left[\left(\bk_3-\frac{z_3}{z_1}\bp\right)\des\right]\left[\left(\bk_2-\frac{z_3}{z_1}\bk_1\right)\de\right]+ z_3^2\left[\left(\bk_3-\frac{z_3}{z_1}\bp\right)\de\right]\left[\left(\bk_2-\frac{z_3}{z_1}\bk_1\right)\des\right]\Bigg\}\bk_1\de \nonumber \\
&+ \frac{2^4 (l^+)^2 \sqrt{z_2} z_3^2}{\sqrt{z_1}(z_1-z_3)}[\bk_1^2+z_1 z_2Q^2]\left(\bk_3 - \frac{z_3}{z_1}\bp\right)\de,\label{N5}
\end{align}

\begin{align}
N_6^{+;+} = &\frac{2^5 (l^+)^2  z_1^{3/2}\sqrt{z_2}}{(z_2-z_3)^2} \Bigg\{ z_3^2 \left[\left(\bk_3 - \frac{z_3}{z_2}\bq\right)\de\right]\left[\left(\bk_2-\frac{z_3}{z_2}\bk_1\right)\de^*\right] + z_2^2 \left[\left(\bk_3 - \frac{z_3}{z_2}\bq\right)\de^*\right]\left[\left(\bk_2-\frac{z_3}{z_2}\bk_1\right)\de\right] \Bigg\}\bk_1\de \nonumber \\
& - \frac{2^4 (l^+)^2 \sqrt{z_1} z_3^2}{\sqrt{z_2} (z_2-z_3)} [\bk_1^2+z_1 z_2Q^2]\left(\bk_3-\frac{z_3}{z_2}\bq\right)\de,\nonumber \\
N_6^{+;-} =&\frac{-2^5 (l^+)^2  z_2^{3/2}\sqrt{z_1}}{(z_2-z_3)^2} \Bigg\{ z_3^2 \left[\left(\bk_3 - \frac{z_3}{z_2}\bq\right)\des\right]\left[\left(\bk_2-\frac{z_3}{z_2}\bk_1\right)\de\right] + z_2^2 \left[\left(\bk_3 - \frac{z_3}{z_2}\bq\right)\de\right]\left[\left(\bk_2-\frac{z_3}{z_2}\bk_1\right)\des\right] \Bigg\}\bk_1\de
\end{align}

\begin{align}
N_{7}^{+;+} =& \frac{-2^5 (l^+)^2  z_3 \sqrt{z_1z_2}}{(1-z_3)(z_1-z_3)^2}\Bigg\{ z_1 z_2 \left[\left( \bk_3 - \frac{z_3}{z_1} \bp\right)\de\right] \Big[\Big( z_2 \bk_1-(1-z_3)\bk_2 \Big)\de^*\Big]\nonumber \\ & + z_3(1-z_3)\left[\left( \bk_3 - \frac{z_3}{z_1} \bp\right)\de^*\right] \Big[\Big( z_2 \bk_1-(1-z_3)\bk_2 \Big)\de \Big]\Bigg\}\bk_1\de\nonumber \\
& - \frac{2^4 (l^+)^2 (z_1 z_2)^{3/2}}{(z_1-z_3)(1-z_3)}\Big[ \bk_1^2 + z_3(1-z_3)Q^2\Big] \left[\left( \bk_3 - \frac{z_3}{z_1} \bp\right)\de\right], \nonumber\\
N_{7}^{+;-} =& \frac{2^5 (l^+)^2   \sqrt{z_1z_2}}{(z_1-z_3)^2}\Bigg\{ z_1 z_2 \left[\left( \bk_3 - \frac{z_3}{z_1} \bp\right)\des\right] \Big[\Big( z_2 \bk_1-(1-z_3)\bk_2 \Big)\de\Big]\nonumber \\ & + z_3(1-z_3)\left[\left( \bk_3 - \frac{z_3}{z_1} \bp\right)\de\right] \Big[\Big( z_2 \bk_1-(1-z_3)\bk_2 \Big)\des \Big]\Bigg\}\bk_1\de,
\end{align}

\begin{align}
N_{8}^{+;+} = &\frac{-2^5 (l^+)^2  \sqrt{z_1z_2}}{ (z_2 - z_3)^2} \Bigg\{ z_3(1-z_3)\left[\left( \bk_3 - \frac{z_3}{z_2} \bq\right)\cdot \be\right]\Big[\Big(z_1 \bk_1 - (1-z_3) \bk_2\Big)\cdot \be^*\Big] \nonumber \\
&+ z_1z_2\left[\left( \bk_3 - \frac{z_3}{z_2} \bq\right)\cdot \be^*\right]\Big[\Big(z_1 \bk_1 - (1-z_3) \bk_2\Big)\cdot \be\Big]     \Bigg\}\bk_1\de,\nonumber \\
N_{8}^{+;-} = &\frac{2^5 (l^+)^2z_3  \sqrt{z_1z_2}}{ (1-z_3)(z_2 - z_3)^2} \Bigg\{ z_3(1-z_3)\left[\left( \bk_3 - \frac{z_3}{z_2} \bq\right)\cdot \be^*\right]\Big[\Big(z_1 \bk_1 - (1-z_3) \bk_2\Big)\cdot \be\Big] \nonumber \\
&+ z_1z_2\left[\left( \bk_3 - \frac{z_3}{z_2} \bq\right)\cdot \be\right]\Big[\Big(z_1 \bk_1 - (1-z_3) \bk_2\Big)\cdot \be^*\Big]     \Bigg\}\bk_1\de.\nonumber \\
& + \frac{2^4 (l^+)^2(z_1z_2)^{3/2}}{(z_2-z_3)(1-z_3)} \Big[\bk_1^2 + z_3(1-z_3)Q^2\Big]\left[\left(\bk_3-\frac{z_3}{z_2}\bq\right)\de\right],
\end{align}

\begin{align}
N_9^{+;+} = &2^4  (l^+)^2 z_1^{3/2}\sqrt{z_2}\left[k_2^2 +\frac{(2z_1-z)}{z} (k_2-p)^2 - \frac{[z_1^2+(z_1-z)^2]}{z_1 z} p^2\right]\bk_1\de,\nonumber\\
N_9^{+;+} = &-2^4  (l^+)^2 z_2^{3/2}\sqrt{z_1}\left[k_2^2 +\frac{(2z_1-z)}{z} (k_2-p)^2 - \frac{[z_1^2+(z_1-z)^2]}{z_1 z} p^2\right]\bk_1\de,
\end{align}

\begin{align}
N_{10}^{+;+} =&-2^4 (l^+)^2 z_1^{3/2}\sqrt{z_2} \left[ k_2^2 + \frac{(2z_2-z)}{z} (k_2-q)^2 - \frac{[z_2^2+(z_2-z)^2]}{z_2 z} q^2\right]\bk_1\de,\nonumber \\
N_{10}^{+;-} =&2^4 (l^+)^2 z_2^{3/2}\sqrt{z_1} \left[ k_2^2 + \frac{(2z_2-z)}{z} (k_2-q)^2 - \frac{[z_2^2+(z_2-z)^2]}{z_2 z} q^2\right]\bk_1\de,
\end{align}

\begin{align}
N_{11}^{+;+} =& 2^4 (l^+)^2  z_1^{3/2}\sqrt{z_2} \left[ \frac{\left[ (k_2^+)^2 + (p^+)^2\right]}{p^+(k_2^+-p^+)} k_1^2 + \frac{(p^++k_2^+)}{(p^+-k_2^+)} k_2^2 + (k_1-k_2)^2\right]\bk_1\de,\nonumber\\
N_{11}^{+;-} =&- 2^4 (l^+)^2  z_2^{3/2}\sqrt{z_1} \left[ \frac{\left[ (k_2^+)^2 + (p^+)^2\right]}{p^+(k_2^+-p^+)} k_1^2 + \frac{(p^++k_2^+)}{(p^+-k_2^+)} k_2^2 + (k_1-k_2)^2\right]\bk_1\de \nonumber \\
&- \frac{2^4 z_2^{3/2} z_3 (l^+)^2}{\sqrt{z_1}(z_1-z_3)}k_1^2(z_3\bk_1-z_1\bk_2)\de,
\end{align}

\begin{align}
N_{12}^{+;+} =& 2^4 (l^+)^2  z_1^{3/2}\sqrt{z_2}\left[ \frac{\left[ (k_2^+)^2 + (q^+)^2\right]}{q^+(k_2^+-q^+)} k_1^2 + \frac{(q^++k_2^+)}{(q^+-k_2^+)} k_2^2 + (k_1-k_2)^2\right]\bk_1\de \nonumber \\
&+ \frac{2^4 (l^+)^2 z_1^{3/2} z_3 }{\sqrt{z_2}(z_2-z_3)} k_1^2 (z_3\bk_1-z_2\bk_2)\de\nonumber ,\\
N_{12}^{+;+} =&- 2^4 (l^+)^2  z_2^{3/2}\sqrt{z_1}\left[ \frac{\left[ (k_2^+)^2 + (q^+)^2\right]}{q^+(k_2^+-q^+)} k_1^2 + \frac{(q^++k_2^+)}{(q^+-k_2^+)} k_2^2 + (k_1-k_2)^2\right]\bk_1\de,
\end{align}

\begin{align}
&N_{13}^{+;+} = 2^5  (l^+)^2 (z_1+z)\sqrt{z_1 z_2}\Bigg[ z_1z\left(\frac{\bp\de}{z_1}-\frac{\bq\de}{z_2}\right)\left(\frac{\bp\des}{z_1}-\frac{\bk_2\des}{z}\right)\nonumber \\
& +z^2\left(\frac{\bk_2\de}{z}-\frac{\bq\de}{z_2}\right)\left(\frac{\bp\des}{z_1}-\frac{\bk_2\des}{z}\right)  -z_2z\left(\frac{\bk_2\de}{z}-\frac{\bq\de}{z_2}\right)\left(\frac{\bp\des}{z_1}-\frac{\bq\des}{z_2}\right)\nonumber \\
& - p\cdot q -\frac{(z_1+z)}{2z}(k_2-q)^2 + \frac{(z_2-z)}{2z}(k_2+p)^2\Bigg]\bk_1\de,\nonumber \\
&N_{13}^{+;-} = -2^5  (l^+)^2 (z_2-z)\sqrt{z_1 z_2}\Bigg[ z_1z\left(\frac{\bp\des}{z_1}-\frac{\bq\des}{z_2}\right)\left(\frac{\bp\de}{z_1}-\frac{\bk_2\de}{z}\right)\nonumber \\
& +z^2\left(\frac{\bk_2\des}{z}-\frac{\bq\des}{z_2}\right)\left(\frac{\bp\de}{z_1}-\frac{\bk_2\de}{z}\right)  -z_2z\left(\frac{\bk_2\des}{z}-\frac{\bq\des}{z_2}\right)\left(\frac{\bp\de}{z_1}-\frac{\bq\de}{z_2}\right)\nonumber \\
& - p\cdot q -\frac{(z_1+z)}{2z}(k_2-q)^2 + \frac{(z_2-z)}{2z}(k_2+p)^2\Bigg]\bk_1\de,
\end{align}

\begin{align}
N_{14(1)}^{+;+} =& \frac{-2^5 (l^+)^2 \sqrt{z_1 z_2}}{(1-z_3)(z_1-z_3)}\Bigg[ \frac{z_1 z_2}{2}\Big[\bk_1^2+z_3(1-z_3)Q^2\Big]+\frac{z_3(1-z_3)}{2}\Big[\bk_2^2+z_1z_2Q^2\Big]  \nonumber \\
&+(z_1-z_3)\Big[\Big(z_2\bk_1+z_3\bk_2\Big)\de\Big]\Big[\Big(z_1\bk_1+(1-z_3)\bk_2\Big)\des\Big]\Bigg]\bk_1\de \nonumber  \\
&+ \frac{2^4 (l^+)^2 (z_1 z_2)^{3/2}}{z_3(z_1-z_3)(1-z_3)} [\bk_1^2+z_3(1-z_3)Q^2](z_1\bk_1-z_3\bk_2)\de,\nonumber \\
N_{14(1)}^{+;-} =& \frac{2^5 (l^+)^2 \sqrt{z_1 z_2}}{z_3(z_1-z_3)}\Bigg[ \frac{z_1 z_2}{2}\Big[\bk_1^2+z_3(1-z_3)Q^2\Big]+\frac{z_3(1-z_3)}{2}\Big[\bk_2^2+z_1z_2Q^2\Big]  \nonumber \\
&+(z_1-z_3)\Big[\Big(z_2\bk_1+z_3\bk_2\Big)\des\Big]\Big[\Big(z_1\bk_1+(1-z_3)\bk_2\Big)\de\Big]\Bigg]\bk_1\de, \nonumber  \\
N_{14(2)}^{+;+} =&  \frac{-2^5 (l^+)^2 \sqrt{z_1 z_2}}{(1-z_3)(z_3-z_1)}\Bigg[ \frac{z_1 z_2}{2}\Big[\bk_1^2+z_3(1-z_3)Q^2\Big]+\frac{z_3(1-z_3)}{2}\Big[\bk_2^2+z_1z_2Q^2\Big]  \nonumber \\
&+(z_3-z_1)\Big[\Big(z_2\bk_1+z_3\bk_2\Big)\de\Big]\Big[\Big(z_1\bk_1+(1-z_3)\bk_2\Big)\des\Big]\Bigg]\bk_1\de,\\
N_{14(2)}^{+;-} =&  \frac{2^5 (l^+)^2 \sqrt{z_1 z_2}}{z_3(z_3-z_1)}\Bigg[ \frac{z_1 z_2}{2}\Big[\bk_1^2+z_3(1-z_3)Q^2\Big]+\frac{z_3(1-z_3)}{2}\Big[\bk_2^2+z_1z_2Q^2\Big]  \nonumber \\
&+(z_3-z_1)\Big[\Big(z_2\bk_1+z_3\bk_2\Big)\des\Big]\Big[\Big(z_1\bk_1+(1-z_3)\bk_2\Big)\de\Big]\Bigg]\bk_1\de \nonumber \\
& + \frac{2^4 (l^+)^2 (z_1 z_2)^{3/2} }{z_3(z_1-z_3)(1-z_3)} [\bk_1^2 +z_3(1-z_3)Q^2](z_2\bk_1-(1-z_3)\bk_2)\de. \label{N14}
\end{align}

In all these expressions, the momentum fractions $z$ and $z_3$ are all defined in the same was as in~\cite{Bergabo:2022tcu}.




\section{Results}\label{Production cross section}

To calculate the $\mathcal{O}(\alpha_s)$ corrections to the production cross section we need to multiply the helicity amplitudes with the corresponding conjugate amplitudes. We'll write the real corrections as $\sigma_{i\times j}$ for $i,j=1,...,4$ and the virtual corrections as $\sigma_i$ for $i=5,...,14$. The details are shown in~\cite{Bergabo:2022tcu} and here we just show the final results. The $T$ label signifies that we are including contributions only from transversely polarized photons, and imply that we have summed over all outgoing polarizations. 
Furthermore and for the sake of brevity here we omit a factor of $\delta(1-z_1-z_2-z)$ in the real corrections and $\delta(1-z_1-z_2)$ in the virtual corrections and restore them at the end. In many cases, it is easiest to write the results in coordinate space with the radiation kernel $\Delta^{(3)}_{ij}$ defined as follows.

\begin{align}
\Delta^{(3)}_{ij} = \frac{\bx_{3i}\cdot\bx_{3j}}{\bx_{3i}^2\bx_{3j}^2}.
\end{align}

\noindent The next to leading order corrections are then, 

{\small 
\begin{align}
&\frac{\dd \sigma_{1\times 1}^T}{\dd^2 \bp \, \dd^2 \bq\, \dd y_1 \, \dd y_2} = \frac{e^2 g^2 Q^2N_c^2 z_2^2 (1-z_2) [z_1^2 z_2^2 + (z_1^2+z_2^2)(1-z_2)^2+ (1-z_2)^4]}{2(2\pi)^{10}z_1}\int \frac{\dd z}{z} \int \dd^{10}\bx [S_{122^\p 1^\p}-S_{12}-S_{1^\p 2^\p} +1] \nonumber \\
& e^{i\bp\cdot\bx_{1^\p 1}}e^{i\bq\cdot\bx_{2^\p 2}}K_1(|\bx_{12}|Q_2)K_1(|\bx_{1^\p 2^\p}|Q_2) \frac{\bx_{12}\cdot\bx_{1^\p 2^\p}}{|\bx_{12}||\bx_{1^\p 2^\p}|}e^{i\frac{z}{z_1}\bp\cdot\bx_{1^\p 1}}\Delta^{(3)}_{1^\p 1}. \\
&\frac{\dd \sigma_{2\times 2}^T}{\dd^2 \bp \, \dd^2 \bq\, \dd y_1 \, \dd y_2} = \frac{e^2 g^2 Q^2 N_c^2 z_1^2 (1-z_1) [z_1^2 z_2^2 + (z_1^2+z_2^2)(1-z_1)^2+ (1-z_1)^4]}{2(2\pi)^{10}z_2}\int\frac{\dd z}{z} \int \dd^{10}\bx[S_{122^\p 1^\p}-S_{12}-S_{1^\p 2^\p} +1]\nonumber \\
&e^{i\bp\cdot\bx_{1^\p 1}}e^{i\bq\cdot\bx_{2^\p 2}} K_1(|\bx_{12}|Q_1)K_1(|\bx_{1^\p 2^\p}|Q_1) \frac{\bx_{12}\cdot\bx_{1^\p 2^\p}}{|\bx_{12}||\bx_{1^\p 2^\p}|} e^{i\frac{z}{z_2}\bq\cdot\bx_{2^\p 1}} \Delta^{(3)}_{2^\p 2}.\\
&\frac{\dd \sigma_{1\times 2}^T}{\dd^2 \bp \, \dd^2 \bq\, \dd y_1 \, \dd y_2} = \frac{e^2g^2Q^2N_c^2\sqrt{z_1z_2(1-z_1)(1-z_2)}}{2(2\pi)^{10}}\int\frac{\dd z}{z} \int\dd^{10} \bx[S_{12}S_{1^\p 2^\p}-S_{12}-S_{1^\p 2^\p} +1] e^{i\bp\cdot\bx_{1^\p 1}} e^{i\bq\cdot\bx_{2^\p 2}}\nonumber \\
&K_1(|\bx_{12}|Q_2)K_1(|\bx_{1^\p 2^\p}|Q_1) 4\Re \Bigg[ \frac{(\bx_{12}\de)(\bx_{1^\p 2^\p}\des)}{|\bx_{12}||\bx_{1^\p2^\p}|}\Bigg\{ (z_1^2+z_2^2)(1-z_1)(1-z_2) \frac{(\bx_{31}\de)(\bx_{2^\p 3}\des)}{\bx_{31}^2 \bx_{2^\p 3}^2} \nonumber \\
&+z_1z_2((1-z_1)^2+(1-z_2)^2)\frac{(\bx_{31}\des)(\bx_{2^\p 3}\de)}{\bx_{31}^2 \bx_{2^\p 3}^2}\Bigg\}\Bigg]e^{i\frac{z}{z_1}\bp\cdot\bx_{31}}e^{i\frac{z}{z_2}\bq\cdot\bx_{2^\p 3}}.\\
&\frac{\dd \sigma_{3\times 3}^{T}}{\dd^2 \bp \, \dd^2 \bq\, \dd y_1 \, \dd y_2}  =\frac{e^2 g^2 Q^2 N_c^2 z_1 z_2^3}{2(2\pi)^{10}}\int\frac{\dd z}{z} \int \dd^{10}\bx  [S_{11^\p} S_{22^\p} - S_{13}S_{23} - S_{1^\p 3} S_{2^\p 3}+1] e^{i\bp\cdot\bx_{1^\p 1}}e^{i\bq\cdot\bx_{2^\p 2}} \nonumber \\
& \frac{K_1(QX)K_1(QX^\p)}{XX^\p} \, 4\Re \Bigg[(z_1^2+z_2^2)\frac{(\bx_{31}\de)(\bx_{3 1^\p}\des)}{\bx_{31}^2\bx_{3 1^\p}^2} [(z_1\bx_{12}+z\bx_{32})\de][(z_1\bx_{1^\p 2^\p}+z\bx_{3 2^\p})\des] \nonumber \\
&+\left((1-z_2)^2 + \frac{(z_1z_2)^2}{(1-z_2)^2}\right) \frac{(\bx_{31}\des)(\bx_{3 1^\p}\de)}{\bx_{31}^2\bx_{3 1^\p}^2}[(z_1\bx_{12}+z\bx_{32})\de][(z_1\bx_{1^\p 2^\p}+z\bx_{3 2^\p})\des] \nonumber \\
&-\frac{z_1^2 z_2 z}{2(1-z_2)^2}\left\{ \frac{(\bx_{31}\des)}{\bx_{31}^2}[(z_1\bx_{12}+z\bx_{32})\de] + \frac{(\bx_{31^\p}\de)}{\bx_{31^\p}^2}[(z_1\bx_{1^\p2^\p}+z\bx_{32^\p})\des]\right\} +\frac{z_1^2 z^2}{4(1-z_2)^2}\Bigg].\\
&\frac{\dd \sigma_{4\times 4}^{T}}{\dd^2 \bp\, \dd^2 \bq\, \dd y_1 \, \dd y_2} =\frac{e^2 g^2 Q^2 N_c^2 z_2 z_1^3}{2(2\pi)^{10}}\int\frac{\dd z}{z} \int \dd^{10}\bx  [S_{11^\p} S_{22^\p} - S_{13}S_{23} - S_{1^\p 3} S_{2^\p 3}+1] e^{i\bp\cdot\bx_{1^\p 1}}e^{i\bq\cdot\bx_{2^\p 2}}\nonumber \\
& \frac{K_1(QX)K_1(QX^\p)}{XX^\p}\, 4\Re \Bigg[(z_1^2+z_2^2)\frac{(\bx_{32}\de)(\bx_{3 2^\p}\des)}{\bx_{32}^2\bx_{3 2^\p}^2} [(z_2\bx_{21}+z\bx_{31})\de][(z_2\bx_{2^\p 1^\p}+z\bx_{3 1^\p})\des] \nonumber \\
&+\left((1-z_1)^2 + \frac{(z_1z_2)^2}{(1-z_1)^2}\right) \frac{(\bx_{32}\des)(\bx_{3 2^\p}\de)}{\bx_{32}^2\bx_{3 2^\p}^2}[(z_2\bx_{21}+z\bx_{31})\de][(z_2\bx_{2^\p 1^\p}+z\bx_{3 1^\p})\des] \nonumber \\
&-\frac{z_2^2 z_1 z}{2(1-z_1)^2}\left\{ \frac{(\bx_{32}\des)}{\bx_{32}^2}[(z_2\bx_{21}+z\bx_{31})\de] + \frac{(\bx_{32^\p}\de)}{\bx_{32^\p}^2}[(z_2\bx_{2^\p1^\p}+z\bx_{31^\p})\des]\right\} +\frac{z_2^2 z^2}{4(1-z_1)^2}\Bigg].\\
&\frac{\dd \sigma_{3\times 4}^T}{\dd^2 \bp\,\dd^2\bq\,\dd y_1\, \dd y_2} = \frac{e^2 g^2 Q^2 N_c^2(z_1 z_2)^2}{2(2\pi)^{10}}\int\frac{\dd z}{z} \int \dd^{10}\bx  [S_{11^\p} S_{22^\p} - S_{13}S_{23} - S_{1^\p 3} S_{2^\p 3}+1]e^{i\bp\cdot\bx_{1^\p 1}}e^{i\bq\cdot\bx_{2^\p 2}} \nonumber \\
& \frac{K_1(QX)K_1(QX^\p)}{XX^\p} \,4\Re\Bigg[(z_1^2+z_2^2)\frac{(\bx_{31}\de)(\bx_{3 2^\p}\des)}{\bx_{31}^2\bx_{3 2^\p}^2} [(z_1\bx_{12}+z\bx_{32})\de][(z_2\bx_{2^\p 1^\p}+z\bx_{3 1^\p})\des] \nonumber \\
&+ \frac{z_1z_2}{(1-z_1)(1-z_2)}[(1-z_1)^2+(1-z_2)^2] \frac{(\bx_{31}\des)(\bx_{3 2^\p}\de)}{\bx_{31}^2\bx_{3 2^\p}^2}[(z_1\bx_{12}+z\bx_{32})\de][(z_2\bx_{2^\p 1^\p}+z\bx_{3 1^\p})\des] \nonumber \\
&-\frac{z_2 z(1-z_2)}{2(1-z_1)} \frac{(\bx_{31}\des)}{\bx_{31}^2}[(z_1\bx_{12}+z\bx_{32})\de]-\frac{z_1z(1-z_1)}{2(1-z_2)}\frac{(\bx_{3 2^\p}\de)}{\bx_{3 2^\p}^2} [(z_2\bx_{2^\p 1^\p}+z\bx_{3 1^\p})\des]\Bigg].
\end{align}

\begin{align}
&\frac{\dd \sigma_{1\times 3}^T}{\dd^2 \bp\, \dd^2 \bq \, \dd y_1 \, \dd y_2} =  \frac{-e^2 g^2 Q^2N_c^2 z_2^{5/2}\sqrt{1-z_2}}{2(2\pi)^{10}}\int \frac{\dd z}{z} \int\dd^{10}\bx  [S_{122^\p3} S_{1^\p3} - S_{1^\p 3} S_{2^\p 3} - S_{12} + 1]  e^{i\bp\cdot\bx_{1^\p 1}}e^{i\bq\cdot\bx_{2^\p 2}}  \nonumber \\
&\frac{K_1(|\bx_{12}|Q_2) K_1(QX^\p)}{X^\p}4\Re \Bigg[ (1-z_2)(z_1^2+z_2^2)\frac{(\bx_{12}\de)(\bx_{3 1^\p}\des)}{|\bx_{12}|\bx_{3 1^\p}^2} \frac{(\bx_{31}\de)[(z_1\bx_{1^\p 2^\p}+z\bx_{3 2^\p})\des]}{\bx_{31}^2} \nonumber \\
&+\left((1-z_2)^3+\frac{(z_1z_2)^2}{1-z_2}\right) \frac{(\bx_{12}\de)(\bx_{3 1^\p}\de)}{|\bx_{12}|\bx_{3 1^\p}^2} \frac{(\bx_{31}\des)[(z_1\bx_{1^\p 2^\p}+z\bx_{3 2^\p})\des]}{\bx_{31}^2} -\frac{z_1^2 z_2 z}{2(1-z_2)} \frac{(\bx_{12}\de)}{|\bx_{12}|} \frac{(\bx_{31}\des)}{\bx_{31}^2}\Bigg]e^{i\frac{z}{z_1}\bp\cdot\bx_{31}}. \\
&\frac{\dd \sigma_{1\times 4}^T}{\dd^2 \bp\, \dd^2 \bq\, \dd y_1 \, \dd y_2} =  \frac{-e^2 g^2 Q^2N_c^2 z_1 z_2^{3/2}\sqrt{1-z_2}}{2(2\pi)^{10}} \int \frac{ \dd z}{z}\int\dd^{10}\bx  [S_{122^\p3} S_{1^\p3} - S_{1^\p 3} S_{2^\p 3} - S_{12} + 1]  e^{i\bp\cdot\bx_{1^\p 1}}e^{i\bq\cdot\bx_{2^\p 2}} \nonumber \\
&\frac{K_1(|\bx_{12}|Q_2) K_1(QX^\p)}{X^\p} 4\Re \Bigg[ (1-z_2)(z_1^2+z_2^2)\frac{(\bx_{12}\de)(\bx_{3 2^\p}\des)}{|\bx_{12}|\bx_{3 2^\p}^2} \frac{(\bx_{31}\de) [(z_2\bx_{2^\p 1^\p}+z \bx_{3 1^\p})\des]}{\bx_{31}^2} \nonumber \\
&+\frac{z_1z_2}{1-z_1}\left((1-z_1)^2+(1-z_2)^2\right)\frac{(\bx_{12}\de)(\bx_{3 2^\p}\de)}{|\bx_{12}|\bx_{3 2^\p}^2} \frac{(\bx_{31}\des) [(z_2\bx_{2^\p 1^\p}+z \bx_{3 1^\p})\des]}{\bx_{31}^2} -\frac{z_2 z(1-z_2)^2}{2(1-z_1)} \frac{(\bx_{12}\de)}{|\bx_{12}|} \frac{(\bx_{31}\des)}{\bx_{31}^2}\Bigg]e^{i\frac{z}{z_1}\bp\cdot\bx_{31}}. \\
&\frac{\dd \sigma_{2\times 3}^T}{\dd^2 \bp\, \dd^2\bq\, \dd y_1 \, \dd y_2} =  \frac{e^2 g^2 Q^2N_c^2 z_2 z_1^{3/2}\sqrt{1-z_1}}{2(2\pi)^{10}} \int\frac{\dd z}{z}\int\dd^{10}\bx [S_{1231^\p} S_{2^\p 3} - S_{1^\p 3} S_{2^\p 3} - S_{12} + 1] e^{i\bp\cdot\bx_{1^\p 1}}e^{i\bq\cdot\bx_{2^\p 2}}  \nonumber \\
&\frac{K_1(|\bx_{12}|Q_1) K_1(QX^\p)}{X^\p}4\Re \Bigg[ (1-z_1)(z_1^2+z_2^2)\frac{(\bx_{12}\de)(\bx_{3 1^\p}\des)}{|\bx_{12}|\bx_{3 1^\p}^2} \frac{(\bx_{32}\de) [(z_1\bx_{1^\p 2^\p}+z \bx_{3 2^\p})\des]}{\bx_{32}^2} \nonumber \\
&+\frac{z_1z_2}{1-z_2}\left((1-z_1)^2+(1-z_2)^2\right)\frac{(\bx_{12}\de)(\bx_{3 1^\p}\de)}{|\bx_{12}|\bx_{3 1^\p}^2} \frac{(\bx_{32}\des) [(z_1\bx_{1^\p 2^\p}+z \bx_{3 2^\p})\des]}{\bx_{32}^2} -\frac{z_1 z(1-z_1)^2}{2(1-z_2)} \frac{(\bx_{12}\de)}{|\bx_{12}|} \frac{(\bx_{32}\des)}{\bx_{32}^2}\Bigg]e^{i\frac{z}{z_2}\bq\cdot\bx_{32}}. \\
&\frac{\dd \sigma_{2\times 4}^T}{\dd^2 \bp\, \dd^2 \bq\, \dd y_1 \, \dd y_2} =  \frac{e^2 g^2 Q^2N_c^2 z_1^{5/2}\sqrt{1-z_1}}{2(2\pi)^{10}} \int \frac{\dd z}{z}\int\dd^{10}\bx [S_{1231^\p} S_{2^\p 3} - S_{1^\p 3} S_{2^\p 3} - S_{12} + 1] e^{i\bp\cdot\bx_{1^\p 1}}e^{i\bq\cdot\bx_{2^\p 2}}  \nonumber \\
&\frac{K_1(|\bx_{12}|Q_1) K_1(QX^\p)}{X^\p}4\Re \Bigg[ (1-z_1)(z_1^2+z_2^2)\frac{(\bx_{12}\de)(\bx_{3 2^\p}\des)}{|\bx_{12}|\bx_{3 2^\p}^2} \frac{(\bx_{32}\de)[(z_2\bx_{2^\p 1^\p}+z\bx_{3 1^\p})\des]}{\bx_{32}^2} \nonumber \\
&+\left((1-z_1)^3+\frac{(z_1z_2)^2}{1-z_1}\right) \frac{(\bx_{12}\de)(\bx_{3 2^\p}\de)}{|\bx_{12}|\bx_{3 2^\p}^2} \frac{(\bx_{32}\des)[(z_2\bx_{2^\p 1^\p}+z\bx_{3 1^\p})\des]}{\bx_{32}^2} -\frac{z_2^2 z_1 z}{2(1-z_1)} \frac{(\bx_{12}\de)}{|\bx_{12}|} \frac{(\bx_{32}\des)}{\bx_{32}^2}\Bigg]e^{i\frac{z}{z_2}\bq\cdot\bx_{32}}. 
\end{align}
}

\begin{align}
&\frac{\dd \sigma^{T}_5}{\dd^2 \bp \, \dd^2 \bq \, \dd y_1 \, \dd y_2} =  \frac{e^2g^2Q^2 N_c^2z_2^{5/2} \sqrt{z_1}}{2(2\pi)^{10} } \int_0^{z_1} \frac{\dd z}{z} \dd^{10} \bx \,[S_{322^\p 1^\p} S_{13} - S_{13}S_{23} - S_{1^\p 2^\p} + 1] \nonumber \\
&\frac{K_1(QX_5)K_1(|\bx_{1^\p 2^\p}|Q_1)}{X_5\bx_{31}^2 |\bx_{1^\p 2^\p}|} e^{i\bp\cdot(\bx_1^\p-\bx_1)}e^{i\bq\cdot(\bx_2^\p-\bx_2)} e^{-i\frac{z}{z_1} \bp\cdot(\bx_3-\bx_1)} \nonumber \\
& \bx_{1^\p 2^\p}\cdot\left[(z_1^2+z_2^2)(z_1^2+(z_1-z)^2)\bx_{32} +(z_1-z)\left(z_1(z_1^2+(z_1-z)^2)+z_2[z_1z_2+(z_1-z)(z_2+z)]\right)\bx_{13}\right] .
\end{align}

\begin{align}
&\frac{\dd \sigma^{T}_6}{\dd^2 \bp \, \dd^2 \bq \, \dd y_1 \, \dd y_2} = -\frac{e^2g^2Q^2N_c^2z_1^{5/2} \sqrt{z_2}}{2(2\pi)^{10}} \int_0^{z_2} \frac{\dd z}{z} \dd^{10} \bx [S_{132^\p 1^\p}S_{23} - S_{13}S_{23} - S_{1^\p 2^\p}+1] \nonumber \\
&\frac{K_1(QX_6)K_1(|\bx_{1^\p 2^\p}|)}{X_6 |\bx_{1^\p 2^\p}|\bx_{32}^2}e^{i\bp\cdot(\bx_1^\p-\bx_1)}e^{i\bq\cdot(\bx_2^\p-\bx_2)}e^{-i\frac{z}{z_2}\bq\cdot(\bx_3-\bx_2)}\nonumber \\
&\bx_{1^\p 2^\p}\cdot\left[(z_1^2+z_2^2)(z_2^2+(z_2-z)^2)\bx_{31}+(z_2-z)[z_2(z_2^2+(z_2-z)^2)+z_1(z_1z_2+(z_2-z)(z_1+z))]\bx_{23}\right].
\end{align}

\begin{align}
&\frac{\dd \sigma^{T}_{7}}{\dd^2 \bp \, \dd^2 \bq \, \dd y_1 \, \dd y_2} = \frac{e^2g^2Q^2 N_c^2 (z_1z_2)^{3/2}}{2(2\pi)^{10}} \int_0^{z_1}\frac{\dd z \, (z_1-z)}{z} \dd^{10} \bx\,  [S_{322^\p 1^\p} S_{13} - S_{13}S_{23} - S_{1^\p 2^\p} + 1] \frac{ K_1(QX_5)  K_1(|\bx_{1^\p2^\p}|Q_1)}{X_5 \bx_{31}^2 |\bx_{1^\p 2^\p}|} \nonumber \\
&\Bigg[ \frac{4\Re}{\bx_{32}^2}\Bigg\{ (\bx_{1^\p 2^\p}\des)\left[\left(\bx_{31}+\frac{z_2}{z_2+z}\bx_{23}\right)\de\right]\Bigg[\frac{z_2 (z_1-z)}{z_1}\left(z_1^2+(z_2+z)^2\right)(\bx_{31}\de)(\bx_{32}\des) \nonumber \\
&+ (z_2+z)(z_2^2+(z_1-z)^2)(\bx_{32}\de)(\bx_{31}\des)\Bigg]\Bigg\} -\frac{z_1 z_2z}{z_2+z}\bx_{31}\cdot\bx_{1^\p 2^\p}\Bigg]  e^{i\bp\cdot(\bx_1^\p-\bx_1)}e^{i\bq\cdot(\bx_2^\p-\bx_2)} e^{-i\frac{z}{z_1}\bp\cdot(\bx_3-\bx_1)}.
\end{align}

\begin{align}
&\frac{\dd \sigma^{T}_{8}}{\dd^2 \bp \, \dd^2 \bq \, \dd y_1 \, \dd y_2} =  \frac{-e^2g^2 Q^2 N_c^2(z_1 z_2)^{3/2}}{2(2\pi)^{10}} \int_0^{z_2} \frac{\dd z \, (z_2-z)}{z} \dd^{10}\bx\,  [S_{132^\p 1^\p}S_{23} - S_{13}S_{23} - S_{1^\p 2^\p}+1]\frac{K_1(QX_6)K_1(|\bx_{1^\p2^\p}|Q_1)}{X_6 |\bx_{1^\p 2^\p}| \bx_{32}^2} \nonumber \\
&\Bigg[ \frac{4\Re}{\bx_{31}^2} \Bigg\{(\bx_{1^\p 2^\p}\des)\left[\left(\bx_{32}+\frac{z_1}{z_1+z}\bx_{13}\right)\de\right]\Bigg[ (z_1+z)(z_1^2+(z_2-z)^2)(\bx_{31}\de)(\bx_{32}\des) \nonumber \\
&+ \frac{z_1 (z_2-z)}{z_2}\left(z_2^2+(z_1+z)^2\right)(\bx_{32}\de)(\bx_{31}\des)\Bigg]\Bigg\}-\frac{z_1z_2z}{z_1+z}\bx_{32}\cdot\bx_{1^\p 2^\p}\Bigg]e^{i\bp\cdot(\bx_1^\p-\bx_1)}e^{i\bq\cdot(\bx_2^\p-\bx_2)}e^{-i\frac{z}{z_2}\bq\cdot(\bx_3-\bx_2)}.
\end{align}

\begin{align}
\frac{\dd \sigma_9^{T}}{\dd^2 \bp \, \dd^2 \bq\, \dd y_1\, \dd y_2} = &\frac{-e^2 g^2 Q^2 N_c^2 (z_1 z_2)^2(z_1^2+z_2^2)}{4(2\pi)^8} \int \dd^8 \bx \big[S_{122^\p 1^\p} - S_{12} - S_{1^\p 2^\p} + 1\big]\frac{\bx_{12}\cdot\bx_{1^\p 2^\p}}{|\bx_{12}||\bx_{1^\p 2^\p}|} K_1(|\bx_{12}|Q_1) K_1(|\bx_{1^\p 2^\p}|Q_1)  \nonumber \\
&\times e^{i\bp\cdot(\bx_1^\p-\bx_1)}e^{i\bq\cdot(\bx_2^\p-\bx_2)}\int_0^{z_1} \frac{\dd z}{z}\left[ \frac{z_1^2+(z_1-z)^2}{z_1^2}\right] \int \dtwo{\bk} \frac{1}{\left(\bk-\frac{z}{z_1}\bp\right)^2}.
\end{align}

\begin{align}
\frac{\dd \sigma_{10}^{T}}{\dd^2 \bp \, \dd^2 \bq\, \dd y_1\, \dd y_2} = &\frac{-e^2 g^2 Q^2 N_c^2 (z_1 z_2)^2(z_1^2+z_2^2)}{4(2\pi)^8} \int \dd^8 \bx \big[S_{122^\p 1^\p} - S_{12} - S_{1^\p 2^\p} + 1\big]\frac{\bx_{12}\cdot\bx_{1^\p 2^\p}}{|\bx_{12}||\bx_{1^\p 2^\p}|} K_1(|\bx_{12}|Q_1) K_1(|\bx_{1^\p 2^\p}|Q_1)  \nonumber \\
&\times e^{i\bp\cdot(\bx_1^\p-\bx_1)}e^{i\bq\cdot(\bx_2^\p-\bx_2)}\int_0^{z_2} \frac{\dd z}{z}\left[ \frac{z_2^2+(z_2-z)^2}{z_2^2}\right] \int \dtwo{\bk} \frac{1}{\left(\bk-\frac{z}{z_2}\bq\right)^2}.
\end{align}

\begin{align}
&\frac{\dd \sigma_{11}^{T}}{\dd^2 \bp\, \dd^2 \bq\, \dd y_1 \, \dd y_2} = \frac{ie^2g^2QN_c^2z_2^{3/2}\sqrt{z_1}(z_1^2+z_2^2)}{2(2\pi)^7} \int \dd^8 \bx\big[S_{122^\p1^\p}-S_{12}-S_{1^\p 2^\p}+1\big] \frac{K_1(|\bx_{1^\p 2^\p}|Q_1)}{|\bx_{1^\p 2^\p}|} e^{i\bp\cdot(\bx_1^\p-\bx_1)}e^{i\bq\cdot(\bx_2^\p-\bx_2)} \nonumber\\
&\int_0^{z_1} \frac{\dd z}{z^2} \frac{[(z_1-z)^2+z_1^2]}{(z_1-z)} \int \dtwo{\bk_2} \int \dtwo{\bk_1}  \frac{\bk_1\cdot\bx_{1^\p 2^\p}\,e^{i\bk_1\cdot(\bx_1-\bx_2)}}{\big[\bk_1^2+Q_1^2\big]\left[Q^2 +\frac{\bk_1^2}{z_1z_2} +\frac{z_1}{z(z_1-z)}\bk_2^2\right]}
\end{align}

\begin{align}
&\frac{\dd \sigma_{12}^{T}}{\dd^2 \bp\, \dd^2 \bq\, \dd y_1 \, \dd y_2} = \frac{-ie^2g^2QN_c^2z_1^{3/2}\sqrt{z_2}(z_1^2+z_2^2)}{2(2\pi)^7} \int \dd^8 \bx\big[S_{122^\p1^\p}-S_{12}-S_{1^\p 2^\p}+1\big] \frac{K_1(|\bx_{1^\p 2^\p}|Q_1)}{|\bx_{1^\p 2^\p}|} e^{i\bp\cdot(\bx_1^\p-\bx_1)}e^{i\bq\cdot(\bx_2^\p-\bx_2)} \nonumber\\
&\int_0^{z_2} \frac{\dd z}{z^2} \frac{[(z_2-z)^2+z_2^2]}{(z_2-z)} \int \dtwo{\bk_2} \int \dtwo{\bk_1}  \frac{\bk_1\cdot\bx_{1^\p 2^\p}\,e^{i\bk_1\cdot(\bx_2-\bx_1)}}{\big[\bk_1^2+Q_1^2\big]\left[Q^2 +\frac{\bk_1^2}{z_1z_2} +\frac{z_2}{z(z_2-z)}\bk_2^2\right]}
\end{align}

\begin{align}
&\frac{\dd \sigma_{13(1)}^{T}}{\dd^2 \bp\, \dd^2 \bq\, \dd y_1\, \dd y_2} =  \frac{  e^2g^2Q^2N_c^2(z_1z_2)^{3/2}}{2(2\pi)^8}\int_{0}^{z_2} \dd z \sqrt{(z_1+z)(z_2-z)} \int \dd^8\bx\,[S_{12}S_{1^\p 2^\p} - S_{12} - S_{1^\p 2^\p} +1]  e^{i\bp\cdot\bx_{1^\p 1}}e^{i\bq\cdot\bx_{2^\p 2}}\nonumber \\
& \frac{ K_1\left(|\bx_{12}|Q\sqrt{(z_1+z)(z_2-z)}\right)}{|\bx_{12}|} \frac{K_1(|\bx_{1^\p 2^\p}|Q_1)}{|\bx_{1^\p 2^\p}|} \int \dtwo{\bk}e^{i\bk\cdot\bx_{21}}4\Re\Bigg[ (\bx_{12}\de)(\bx_{1^\p 2^\p}\des)\Bigg\{ \frac{\frac{z_2(z_2-z)[z_1(z_1+z)+z_2(z_2-z)]}{2z}}{(z_2\bk-z\bq)^2} \nonumber \\
&+z_1z_2 z\left(\frac{ z_1z\left(\frac{\bp\de}{z_1}-\frac{\bq\de}{z_2}\right)\left(\frac{\bp\des}{z_1}-\frac{\bk\des}{z}\right) +z^2\left(\frac{\bk\de}{z}-\frac{\bq\de}{z_2}\right)\left(\frac{\bp\des}{z_1}-\frac{\bk\des}{z}\right) -z_2z\left(\frac{\bk\de}{z}-\frac{\bq\de}{z_2}\right)\left(\frac{\bp\des}{z_1}-\frac{\bq\des}{z_2}\right) - p\cdot q}{(z_2\bk-z\bq)^2\left[\frac{(z_1\bk-z\bp)^2}{z_1(z_1+z)}-\frac{(z_2\bk-z\bq)^2}{z_2 (z_2-z)}\right]}\right) \nonumber \\
&+z_2^2 z (z_2-z)\left(\frac{ z_1z\left(\frac{\bp\des}{z_1}-\frac{\bq\des}{z_2}\right)\left(\frac{\bp\de}{z_1}-\frac{\bk\de}{z}\right) +z^2\left(\frac{\bk\des}{z}-\frac{\bq\des}{z_2}\right)\left(\frac{\bp\de}{z_1}-\frac{\bk\de}{z}\right) -z_2z\left(\frac{\bk\des}{z}-\frac{\bq\des}{z_2}\right)\left(\frac{\bp\de}{z_1}-\frac{\bq\de}{z_2}\right) - p\cdot q}{(z_1+z)(z_2\bk-z\bq)^2\left[\frac{(z_1\bk-z\bp)^2}{z_1(z_1+z)}-\frac{(z_2\bk-z\bq)^2}{z_2 (z_2-z)}\right]}\right)\Bigg\}\Bigg]. 
\end{align}

\begin{align}
&\frac{\dd \sigma_{13(2)}^{T}}{\dd^2 \bp\, \dd^2 \bq\, \dd y_1\, \dd y_2} =  \frac{  e^2g^2Q^2N_c^2(z_1z_2)^{3/2}}{2(2\pi)^8}\int_{0}^{z_1} \dd z \sqrt{(z_1-z)(z_2+z)} \int \dd^8\bx\,[S_{12}S_{1^\p 2^\p} - S_{12} - S_{1^\p 2^\p} +1]  e^{i\bp\cdot\bx_{1^\p 1}}e^{i\bq\cdot\bx_{2^\p 2}}\nonumber \\
& \frac{ K_1\left(|\bx_{12}|Q\sqrt{(z_1-z)(z_2+z)}\right)}{|\bx_{12}|} \frac{K_1(|\bx_{1^\p 2^\p}|Q_1)}{|\bx_{1^\p 2^\p}|} \int \dtwo{\bk}e^{i\bk\cdot\bx_{12}}4\Re\Bigg[  (\bx_{12}\de)(\bx_{1^\p 2^\p}\des) \Bigg\{ \frac{ \frac{z_1(z_1-z)[z_1(z_1-z)+z_2(z_2+z)]}{2z}}{(z_1\bk-z\bp)^2}     \nonumber \\
&+z_1z_2z\left(\frac{ z_1z\left(\frac{\bp\des}{z_1}-\frac{\bq\des}{z_2}\right)\left(\frac{\bp\de}{z_1}-\frac{\bk\de}{z}\right) -z^2\left(\frac{\bk\des}{z}-\frac{\bq\des}{z_2}\right)\left(\frac{\bp\de}{z_1}-\frac{\bk\de}{z}\right) -z_2z\left(\frac{\bk\des}{z}-\frac{\bq\des}{z_2}\right)\left(\frac{\bp\de}{z_1}-\frac{\bq\de}{z_2}\right) + p\cdot q}{(z_1\bk-z\bp)^2\left[\frac{(z_1\bk-z\bp)^2}{z_1(z_1-z)}-\frac{(z_2\bk-z\bq)^2}{z_2 (z_2+z)}\right]}\right) \nonumber \\
&+z_1^2 z(z_1-z)\left(\frac{ z_1z\left(\frac{\bp\de}{z_1}-\frac{\bq\de}{z_2}\right)\left(\frac{\bp\des}{z_1}-\frac{\bk\des}{z}\right) -z^2\left(\frac{\bk\de}{z}-\frac{\bq\de}{z_2}\right)\left(\frac{\bp\des}{z_1}-\frac{\bk\des}{z}\right) -z_2z\left(\frac{\bk\de}{z}-\frac{\bq\de}{z_2}\right)\left(\frac{\bp\des}{z_1}-\frac{\bq\des}{z_2}\right) + p\cdot q}{(z_2+z)(z_1\bk-z\bp)^2\left[\frac{(z_1\bk-z\bp)^2}{z_1(z_1-z)}-\frac{(z_2\bk-z\bq)^2}{z_2 (z_2+z)}\right]}\right)\Bigg\}\Bigg].
\end{align}
\eq{
&\frac{ \dd \sigma_{14(1)}^{T}}{\dd^2 \bp \, \dd^2 \bq \, \dd y_1 \, \dd y_2} = \frac{-ie^2g^2QN_c^2 (z_1z_2)^{3/2}}{2(2\pi)^7} \int_0^{z_1} \frac{\dd z}{z}\dd^8 \bx\frac{ K_1(|\bx_{1^\p 2^\p}|Q_1)}{|\bx_{1^\p 2^\p}|}[S_{122^\p 1^\p} -S_{1^\p 2^\p}-S_{12}+1]e^{i(\bp\cdot\bx_{1^\p 1} + \bq\cdot\bx_{2^\p 2})} \nonumber \\
&\int  \dtwo{\bk_1}\dtwo{\bk_2}e^{i\bk_2\cdot\bx_{12}} \Bigg[\frac{\br{z_1(z_1-z)+z_2(z_2+z)-z(1-z)} (\bk_2\cdot\bx_{1^\p 2^\p})    }
{\Big[\bk_2^2 + Q_1^2\Big] \left[\pa{\bk_1-\frac{z_1-z}{z_1}\bk_2}^2+\frac{z(z_1-z)}{z_2 z_1^2} \bk_2^2 + \frac{z}{z_1}(z_1-z)Q^2\right] }\nonumber \\
&+\frac{\frac{(z_1-z)}{z_1}(1+z^2-2z_2(z_1-z))(\bk_1\cdot\bx_{1^\p 2^\p}) }{\Big[\bk_1^2 + (z_1-z)(z_2+z)Q^2\Big] \left[\left(\bk_1-\frac{z_1-z}{z_1}\bk_2\right)^2+\frac{z(z_1-z)}{z_2 z_1^2} \bk_2^2 + \frac{z}{z_1}(z_1-z)Q^2\right]} \nonumber \\
& -\frac{Q^2 \frac{(z_1-z)}{z_1}\br{2z_1z_2z(\bk_1\cdot\bx_{1^\p 2^\p}) + z(z+z_2-z_1)^2(\bk_2\cdot\bx_{1^\p 2^\p})}}{\Big[\bk_1^2 + (z_1-z)(z_2+z)Q^2\Big]\Big[\bk_2^2 + Q_1^2\Big]\left[\left(\bk_1-\frac{z_1-z}{z_1}\bk_2\right)^2+\frac{z(z_1-z)}{z_2 z_1^2} \bk_2^2 + \frac{z}{z_1}(z_1-z)Q^2\right]}
\Bigg]
}

\eq{
&\frac{ \dd \sigma_{14(2)}^{T}}{\dd^2 \bp \, \dd^2 \bq \, \dd y_1 \, \dd y_2} = \frac{-ie^2g^2QN_c^2 (z_1z_2)^{3/2}}{2(2\pi)^7} \int_0^{z_2} \frac{\dd z}{z}\dd^8 \bx\frac{ K_1(|\bx_{1^\p 2^\p}|Q_1)}{|\bx_{1^\p 2^\p}|}[S_{122^\p 1^\p} -S_{1^\p 2^\p}-S_{12}+1]e^{i(\bp\cdot\bx_{1^\p 1} + \bq\cdot\bx_{2^\p 2})} \nonumber \\
&\int  \dtwo{\bk_1}\dtwo{\bk_2}e^{i\bk_2\cdot\bx_{12}} \Bigg[\frac{\br{z_2(z_2-z)+z_1(z_1+z)-z(1-z)} (\bk_2\cdot\bx_{1^\p 2^\p})    }
{\Big[\bk_2^2 + Q_1^2\Big] \left[\pa{\bk_1-\frac{z_2-z}{z_2}\bk_2}^2+\frac{z(z_2-z)}{z_1 z_2^2} \bk_2^2 + \frac{z}{z_2}(z_2-z)Q^2\right] }\nonumber \\
&+\frac{\frac{(z_2-z)}{z_2}(1+z^2-2z_1(z_2-z))(\bk_1\cdot\bx_{1^\p 2^\p}) }{\Big[\bk_1^2 + (z_2-z)(z_1+z)Q^2\Big] \left[\left(\bk_1-\frac{z_2-z}{z_2}\bk_2\right)^2+\frac{z(z_2-z)}{z_1 z_2^2} \bk_2^2 + \frac{z}{z_2}(z_2-z)Q^2\right]} \nonumber \\
& -\frac{Q^2 \frac{(z_2-z)}{z_2}\br{2z_1z_2z(\bk_1\cdot\bx_{1^\p 2^\p}) + z(z+z_1-z_2)^2(\bk_2\cdot\bx_{1^\p 2^\p})}}{\Big[\bk_1^2 + (z_2-z)(z_1+z)Q^2\Big]\Big[\bk_2^2 + Q_1^2\Big]\left[\left(\bk_1-\frac{z_2-z}{z_2}\bk_2\right)^2+\frac{z(z_2-z)}{z_1 z_2^2} \bk_2^2 + \frac{z}{z_2}(z_2-z)Q^2\right]}
\Bigg]
}

\noindent These expressions constitute the full result for the one-loop corrections to inclusive quark anti-quark production cross section with transverse photon exchange. We have written these results all in terms of the dipole and quadrupole functions defined in Eq.~(\ref{dipquad}) in the large $N_c$ limit and ignored all subleading $N_c$ terms.

We have also used the following notation for the coordinate dependence of some of the Bessel functions:

\begin{align}
X &= \sqrt{z_1 z_2 \bx_{12}^2 + z_1 z \bx_{13}^2 + z_2 z \bx_{23}^2},\nonumber \\
X_5 &= \sqrt{z_2(z_1-z)\bx_{12}^2 + z(z_1-z) \bx_{13}^2 + z_2 z\, \bx_{23}^2}, \nonumber \\
X_6 &= \sqrt{z_1(z_2-z)\bx_{12}^2 + z_1 z\,\bx_{13}^2 + z(z_2-z)\bx_{23}^2}.
\end{align}

\noindent Note that when $z \to 0$ these all become $|\bx_{12}|\sqrt{z_1 z_2}$. The primed version $X^\prime$ that appears in some real corrections is the same as $X$ above but with $\bx_1,\bx_2 \to \bx_1^\p,\bx_2^\p$. 




\section{Divergences}
\noindent The above expressions are formal in the sense that they contain divergences that render 
them ill-defined unless regulated. As in the case of longitudinal exchange there are 
$4$ types of divergences:

\noindent $\bullet$ Ultraviolet (UV) divergences when loop momentum $\bk \rightarrow \infty$ or equivalently in coordinate space, when the transverse coordinate of the radiated gluon approaches the transverse coordinate $\bx_{i }$ of either quark or antiquark when integrated, i.e.  $\bx_{3 } \rightarrow \bx_{i }$ such that $|\bx_{3 } - \bx_{i }| \rightarrow 0$. The UV structure of the 
production cross section with transverse photon exchange is identical to that of longitudinal photon exchange so that cancellations are identical, i.e.
\begin{align}
&\left[\dd \sigma_{5}+\dd \sigma_{11}\right]_{\text{UV}} = 0, \nonumber \\
&\left[\dd \sigma_{6}+\dd \sigma_{12} \right]_{\text{UV}} = 0, \nonumber \\
&\left[\dd \sigma_{9}+\dd\sigma_{10}+\dd\sigma_{14(1)}+\dd\sigma_{14(2)}\right]_{\text{UV}}  = 0. \label{uvcancels}
\end{align}
with the rest of the contributions being UV finite.

\noindent $\bullet$ Soft divergences when $k^\mu \rightarrow 0$, which in this context corresponds 
to {\it both} transverse momentum in the loop $\bk$ {\it and} the radiated gluon momentum fraction 
$z$ go to zero simultaneously, $\bk , z \rightarrow 0$. Both the real and virtual corrections contain soft divergences, however all soft divergences cancel between real and virtual corrections as shown below,
\begin{align}
&\left[\dd \sigma_{1\times 1} + 2\,\dd \sigma_{9} \right]_{\text{soft}}= 0, \nonumber \\
&\left[\dd \sigma_{2\times 2} + 2\,\dd \sigma_{10}\right]_{\text{soft}} = 0, \nonumber \\
&\left[\dd \sigma_{1\times 2} + \dd\sigma_{13(1)}+\dd\sigma_{13(2)} \right]_{\text{soft}}= 0, \nonumber \\
&\left[\dd \sigma_{3\times 3} + \dd \sigma_{4\times 4} + 2\,\dd \sigma_{3\times 4}\right]_{\text{soft}} = 0,\nonumber\\
&\left[\dd \sigma_{1\times 3} + \dd \sigma_{1\times 4} \right]_{\text{soft}}= 0,\nonumber\\
&\left[\dd \sigma_{2\times 3} + \dd \sigma_{2\times 4}\right]_{\text{soft}} = 0,\nonumber\\
&\left[\dd \sigma_{5}+\dd\sigma_{7} \right]_{\text{soft}}= 0, \nonumber \\
&\left[\dd \sigma_{6}+\dd\sigma_{8}\right]_{\text{soft}} = 0, \nonumber \\
&\left[\dd \sigma_{11} + \dd \sigma_{14(1)}\right]_{\text{soft}} = 0,  \nonumber \\
&\left[\dd \sigma_{12} + \dd \sigma_{14(2)}\right]_{\text{soft}} = 0 .
\end{align}

\noindent $\bullet$ Collinear divergences when the radiated gluon momentum becomes parallel to either quark or anti-quark momentum at {\it finite} $\bk$ and $z$. They are present in diagrams 
$i \mathcal{A}_1, i \mathcal{A}_2$ (real corrections) and in $i \mathcal{A}_9, i \mathcal{A}_{10}$ 
(virtual corrections). These collinear divergences are absorbed into quark-hadron and antiquark-hadron fragmentation functions which makes the fragmentation functions scale dependent, for example 
\begin{align}
D_{h_1/q}(\zho,\mu^2)= \int_{\zho}^1 \frac{\dd \xi}{\xi} D^0_{h_1/q}\left(\frac{\zho}{\xi}\right) \Bigg[ \delta(1-\xi) +\frac{\alpha_s }{2\pi}P_{qq}(\xi)\log\left(\frac{\mu^2}{\Lambda^2}\right)\Bigg],
\end{align}
 defined using a cutoff scheme or 
\begin{align}
D_{h_1/q}(\zho,\mu^2)= \int_{\zho}^1 \frac{\dd \xi}{\xi} D^0_{h_1/q}\left(\frac{\zho}{\xi}\right) \Bigg[ \delta(1-\xi) +\frac{\alpha_s }{\pi}P_{qq}(\xi) \left( \frac{1}{\epsilon} -\log\left(\pi e^{\gamma_E}\mu |\bx_1^\p-\bx_1|\right)\right)\Bigg],
\end{align}
when using dimensional regularization scheme. We refer the reader to~\cite{Bergabo:2022tcu} for full
details. 

\noindent $\bullet$ Rapidity divergences when the momentum fraction $z$ of the gluon goes to zero while the transverse momentum $\bk$ of the gluon remains finite. These are handled by introducing a longitudinal momentum fraction factorization scale $z_f$ and dividing the $z$ integration into 
two regions: $z > z_f$ and $z < z_f$, 
\begin{align}
\int_0^1 \frac{\dd z}{z} f (z) =  \left\{\int_0^{z_f} \frac{\dd z}{z}  + 
\int_{z_f}^1 \frac{\dd z}{z} \, \right\} \, f (z).
\end{align}
The rapidity divergences are present only in the first term above and lead to evolution (renormalization) the dipoles and quadrupoles according to the BK and JIMWLK evolution equations~\cite{Balitsky:1995ub,Kovchegov:1999yj,Jalilian-Marian:1997qno,Jalilian-Marian:1997jhx,Jalilian-Marian:1997ubg,Jalilian-Marian:1998tzv,Kovner:2000pt,Iancu:2000hn,Ferreiro:2001qy}. The second term contains no 
rapidity divergences, it is completely finite and is part of the next to leading order corrections. 

Our final result for the regulated dihadron production cross section can then be symbolically written as sum of several terms (Eq. \ref{convolutioneq}) as shown below

\begin{align}
\dd \sigma^{\gamma^* A \to h_1 h_2 X}= \dd \sigma_{LO}\otimes \text{JIMWLK} + 
\dd\sigma_{LO}\otimes D_{h_1/q}(\zho , \mu^2) \otimes D_{h_2/\barq}(\zht , \mu^2) + 
\dd\sigma_{NLO}^{\text{finite}} \label{convolutioneq}
\end{align}

\noindent The first term contains the $z$ integration region below $z_f$ where the leading order cross section is evolved with the BK/JIMWLK evolution equations. The second term includes the integration region $z > z_f$ where the leading order cross section is convoluted with the DGLAP evolved fragmentation functions for both quark and antiquark. Finally the last term constitutes all the remaining contributions to the NLO cross section which is finite. Presence of the bare fragmentation functions in the first and last terms is implied.

\vspace{0.2in}
\noindent In summary, we have calculated the one-loop corrections to inclusive quark antiquark production in DIS at small $x$ for transverse photons. We have shown the production cross section factorizes: all divergences that appear at the one-loop level are either canceled or absorbed into JIMWLK evolution of dipoles and quadrupoles, and into DGLAP evolution of parton-hadron fragmentation functions. These results are well suited for further phenomenological studies of angular correlations of the dihadrons produced in DIS at small $x$~\cite{Bergabo:2021woe}.



\vspace{0.2in}
\paragraph{Acknowledgements:}We gratefully acknowledge support from the DOE Office of Nuclear Physics through Grant No. DE-SC0002307 and by PSC-CUNY through grant No. 63158-0051. We would like to thank T. Altinoluk, G. Beuf, R. Boussarie, P. Caucal, L. Dixon, Y. Kovchegov, C. Marquet, Y. Mulian, F. Salazar, M. Tevio, R. Venugopalan, W. Vogelsang and B. Xiao for helpful discussions.


\bibliography{mybib}

\begin{thebibliography}{68}
\expandafter\ifx\csname natexlab\endcsname\relax\def\natexlab#1{#1}\fi
\expandafter\ifx\csname bibnamefont\endcsname\relax
  \def\bibnamefont#1{#1}\fi
\expandafter\ifx\csname bibfnamefont\endcsname\relax
  \def\bibfnamefont#1{#1}\fi
\expandafter\ifx\csname citenamefont\endcsname\relax
  \def\citenamefont#1{#1}\fi
\expandafter\ifx\csname url\endcsname\relax
  \def\url#1{\texttt{#1}}\fi
\expandafter\ifx\csname urlprefix\endcsname\relax\def\urlprefix{URL }\fi
\providecommand{\bibinfo}[2]{#2}
\providecommand{\eprint}[2][]{\url{#2}}

\bibitem[{\citenamefont{Iancu and Venugopalan}(2003)}]{Iancu:2003xm}
\bibinfo{author}{\bibfnamefont{E.}~\bibnamefont{Iancu}} \bibnamefont{and}
  \bibinfo{author}{\bibfnamefont{R.}~\bibnamefont{Venugopalan}},
  \emph{\bibinfo{title}{{The Color glass condensate and high-energy scattering
  in QCD}}} (\bibinfo{year}{2003}), pp. \bibinfo{pages}{249--3363},
  \eprint{hep-ph/0303204}.

\bibitem[{\citenamefont{Jalilian-Marian and
  Kovchegov}(2006)}]{Jalilian-Marian:2005ccm}
\bibinfo{author}{\bibfnamefont{J.}~\bibnamefont{Jalilian-Marian}}
  \bibnamefont{and} \bibinfo{author}{\bibfnamefont{Y.~V.}
  \bibnamefont{Kovchegov}}, \bibinfo{journal}{Prog. Part. Nucl. Phys.}
  \textbf{\bibinfo{volume}{56}}, \bibinfo{pages}{104} (\bibinfo{year}{2006}),
  \eprint{hep-ph/0505052}.

\bibitem[{\citenamefont{Weigert}(2005)}]{Weigert:2005us}
\bibinfo{author}{\bibfnamefont{H.}~\bibnamefont{Weigert}},
  \bibinfo{journal}{Prog. Part. Nucl. Phys.} \textbf{\bibinfo{volume}{55}},
  \bibinfo{pages}{461} (\bibinfo{year}{2005}), \eprint{hep-ph/0501087}.

\bibitem[{\citenamefont{Gelis et~al.}(2010)\citenamefont{Gelis, Iancu,
  Jalilian-Marian, and Venugopalan}}]{Gelis:2010nm}
\bibinfo{author}{\bibfnamefont{F.}~\bibnamefont{Gelis}},
  \bibinfo{author}{\bibfnamefont{E.}~\bibnamefont{Iancu}},
  \bibinfo{author}{\bibfnamefont{J.}~\bibnamefont{Jalilian-Marian}},
  \bibnamefont{and}
  \bibinfo{author}{\bibfnamefont{R.}~\bibnamefont{Venugopalan}},
  \bibinfo{journal}{Ann. Rev. Nucl. Part. Sci.} \textbf{\bibinfo{volume}{60}},
  \bibinfo{pages}{463} (\bibinfo{year}{2010}), \eprint{1002.0333}.

\bibitem[{\citenamefont{Morreale and Salazar}(2021)}]{Morreale:2021pnn}
\bibinfo{author}{\bibfnamefont{A.}~\bibnamefont{Morreale}} \bibnamefont{and}
  \bibinfo{author}{\bibfnamefont{F.}~\bibnamefont{Salazar}},
  \bibinfo{journal}{Universe} \textbf{\bibinfo{volume}{7}},
  \bibinfo{pages}{312} (\bibinfo{year}{2021}), \eprint{2108.08254}.

\bibitem[{\citenamefont{Kovner and Wiedemann}(2001)}]{Kovner:2001vi}
\bibinfo{author}{\bibfnamefont{A.}~\bibnamefont{Kovner}} \bibnamefont{and}
  \bibinfo{author}{\bibfnamefont{U.~A.} \bibnamefont{Wiedemann}},
  \bibinfo{journal}{Phys. Rev. D} \textbf{\bibinfo{volume}{64}},
  \bibinfo{pages}{114002} (\bibinfo{year}{2001}), \eprint{hep-ph/0106240}.

\bibitem[{\citenamefont{Jalilian-Marian and
  Kovchegov}(2004)}]{JalilianMarian:2004da}
\bibinfo{author}{\bibfnamefont{J.}~\bibnamefont{Jalilian-Marian}}
  \bibnamefont{and} \bibinfo{author}{\bibfnamefont{Y.~V.}
  \bibnamefont{Kovchegov}}, \bibinfo{journal}{Phys. Rev. D}
  \textbf{\bibinfo{volume}{70}}, \bibinfo{pages}{114017}
  (\bibinfo{year}{2004}), \bibinfo{note}{[Erratum: Phys.Rev.D 71, 079901
  (2005)]}, \eprint{hep-ph/0405266}.

\bibitem[{\citenamefont{Marquet}(2007)}]{Marquet:2007vb}
\bibinfo{author}{\bibfnamefont{C.}~\bibnamefont{Marquet}},
  \bibinfo{journal}{Nucl. Phys. A} \textbf{\bibinfo{volume}{796}},
  \bibinfo{pages}{41} (\bibinfo{year}{2007}), \eprint{0708.0231}.

\bibitem[{\citenamefont{Albacete and Marquet}(2010)}]{Albacete:2010pg}
\bibinfo{author}{\bibfnamefont{J.~L.} \bibnamefont{Albacete}} \bibnamefont{and}
  \bibinfo{author}{\bibfnamefont{C.}~\bibnamefont{Marquet}},
  \bibinfo{journal}{Phys. Rev. Lett.} \textbf{\bibinfo{volume}{105}},
  \bibinfo{pages}{162301} (\bibinfo{year}{2010}), \eprint{1005.4065}.

\bibitem[{\citenamefont{Stasto et~al.}(2012)\citenamefont{Stasto, Xiao, and
  Yuan}}]{Stasto:2011ru}
\bibinfo{author}{\bibfnamefont{A.}~\bibnamefont{Stasto}},
  \bibinfo{author}{\bibfnamefont{B.-W.} \bibnamefont{Xiao}}, \bibnamefont{and}
  \bibinfo{author}{\bibfnamefont{F.}~\bibnamefont{Yuan}},
  \bibinfo{journal}{Phys. Lett. B} \textbf{\bibinfo{volume}{716}},
  \bibinfo{pages}{430} (\bibinfo{year}{2012}), \eprint{1109.1817}.

\bibitem[{\citenamefont{Lappi and Mantysaari}(2013)}]{Lappi:2012nh}
\bibinfo{author}{\bibfnamefont{T.}~\bibnamefont{Lappi}} \bibnamefont{and}
  \bibinfo{author}{\bibfnamefont{H.}~\bibnamefont{Mantysaari}},
  \bibinfo{journal}{Nucl. Phys. A} \textbf{\bibinfo{volume}{908}},
  \bibinfo{pages}{51} (\bibinfo{year}{2013}), \eprint{1209.2853}.

\bibitem[{\citenamefont{Stasto et~al.}(2018)\citenamefont{Stasto, Wei, Xiao,
  and Yuan}}]{Stasto:2018rci}
\bibinfo{author}{\bibfnamefont{A.}~\bibnamefont{Stasto}},
  \bibinfo{author}{\bibfnamefont{S.-Y.} \bibnamefont{Wei}},
  \bibinfo{author}{\bibfnamefont{B.-W.} \bibnamefont{Xiao}}, \bibnamefont{and}
  \bibinfo{author}{\bibfnamefont{F.}~\bibnamefont{Yuan}},
  \bibinfo{journal}{Phys. Lett. B} \textbf{\bibinfo{volume}{784}},
  \bibinfo{pages}{301} (\bibinfo{year}{2018}), \eprint{1805.05712}.

\bibitem[{\citenamefont{Albacete et~al.}(2019)\citenamefont{Albacete,
  Giacalone, Marquet, and Matas}}]{Albacete:2018ruq}
\bibinfo{author}{\bibfnamefont{J.~L.} \bibnamefont{Albacete}},
  \bibinfo{author}{\bibfnamefont{G.}~\bibnamefont{Giacalone}},
  \bibinfo{author}{\bibfnamefont{C.}~\bibnamefont{Marquet}}, \bibnamefont{and}
  \bibinfo{author}{\bibfnamefont{M.}~\bibnamefont{Matas}},
  \bibinfo{journal}{Phys. Rev. D} \textbf{\bibinfo{volume}{99}},
  \bibinfo{pages}{014002} (\bibinfo{year}{2019}), \eprint{1805.05711}.

\bibitem[{\citenamefont{Boussarie
  et~al.}(2021{\natexlab{a}})\citenamefont{Boussarie, M\"antysaari, Salazar,
  and Schenke}}]{Boussarie:2021lkb}
\bibinfo{author}{\bibfnamefont{R.}~\bibnamefont{Boussarie}},
  \bibinfo{author}{\bibfnamefont{H.}~\bibnamefont{M\"antysaari}},
  \bibinfo{author}{\bibfnamefont{F.}~\bibnamefont{Salazar}}, \bibnamefont{and}
  \bibinfo{author}{\bibfnamefont{B.}~\bibnamefont{Schenke}}
  (\bibinfo{year}{2021}{\natexlab{a}}), \eprint{2106.11301}.

\bibitem[{\citenamefont{Fujii et~al.}(2020)\citenamefont{Fujii, Marquet, and
  Watanabe}}]{Fujii:2020bkl}
\bibinfo{author}{\bibfnamefont{H.}~\bibnamefont{Fujii}},
  \bibinfo{author}{\bibfnamefont{C.}~\bibnamefont{Marquet}}, \bibnamefont{and}
  \bibinfo{author}{\bibfnamefont{K.}~\bibnamefont{Watanabe}},
  \bibinfo{journal}{JHEP} \textbf{\bibinfo{volume}{12}}, \bibinfo{pages}{181}
  (\bibinfo{year}{2020}), \eprint{2006.16279}.

\bibitem[{\citenamefont{Kotko et~al.}(2015)\citenamefont{Kotko, Kutak, Marquet,
  Petreska, Sapeta, and van Hameren}}]{Kotko:2015ura}
\bibinfo{author}{\bibfnamefont{P.}~\bibnamefont{Kotko}},
  \bibinfo{author}{\bibfnamefont{K.}~\bibnamefont{Kutak}},
  \bibinfo{author}{\bibfnamefont{C.}~\bibnamefont{Marquet}},
  \bibinfo{author}{\bibfnamefont{E.}~\bibnamefont{Petreska}},
  \bibinfo{author}{\bibfnamefont{S.}~\bibnamefont{Sapeta}}, \bibnamefont{and}
  \bibinfo{author}{\bibfnamefont{A.}~\bibnamefont{van Hameren}},
  \bibinfo{journal}{JHEP} \textbf{\bibinfo{volume}{09}}, \bibinfo{pages}{106}
  (\bibinfo{year}{2015}), \eprint{1503.03421}.

\bibitem[{\citenamefont{van Hameren et~al.}(2016)\citenamefont{van Hameren,
  Kotko, Kutak, Marquet, Petreska, and Sapeta}}]{vanHameren:2016ftb}
\bibinfo{author}{\bibfnamefont{A.}~\bibnamefont{van Hameren}},
  \bibinfo{author}{\bibfnamefont{P.}~\bibnamefont{Kotko}},
  \bibinfo{author}{\bibfnamefont{K.}~\bibnamefont{Kutak}},
  \bibinfo{author}{\bibfnamefont{C.}~\bibnamefont{Marquet}},
  \bibinfo{author}{\bibfnamefont{E.}~\bibnamefont{Petreska}}, \bibnamefont{and}
  \bibinfo{author}{\bibfnamefont{S.}~\bibnamefont{Sapeta}},
  \bibinfo{journal}{JHEP} \textbf{\bibinfo{volume}{12}}, \bibinfo{pages}{034}
  (\bibinfo{year}{2016}), \bibinfo{note}{[Erratum: JHEP 02, 158 (2019)]},
  \eprint{1607.03121}.

\bibitem[{\citenamefont{Altinoluk et~al.}(2021)\citenamefont{Altinoluk,
  Marquet, and Taels}}]{Altinoluk:2021ygv}
\bibinfo{author}{\bibfnamefont{T.}~\bibnamefont{Altinoluk}},
  \bibinfo{author}{\bibfnamefont{C.}~\bibnamefont{Marquet}}, \bibnamefont{and}
  \bibinfo{author}{\bibfnamefont{P.}~\bibnamefont{Taels}},
  \bibinfo{journal}{JHEP} \textbf{\bibinfo{volume}{06}}, \bibinfo{pages}{085}
  (\bibinfo{year}{2021}), \eprint{2103.14495}.

\bibitem[{\citenamefont{Hatta et~al.}(2021)\citenamefont{Hatta, Xiao, Yuan, and
  Zhou}}]{Hatta:2020bgy}
\bibinfo{author}{\bibfnamefont{Y.}~\bibnamefont{Hatta}},
  \bibinfo{author}{\bibfnamefont{B.-W.} \bibnamefont{Xiao}},
  \bibinfo{author}{\bibfnamefont{F.}~\bibnamefont{Yuan}}, \bibnamefont{and}
  \bibinfo{author}{\bibfnamefont{J.}~\bibnamefont{Zhou}},
  \bibinfo{journal}{Phys. Rev. Lett.} \textbf{\bibinfo{volume}{126}},
  \bibinfo{pages}{142001} (\bibinfo{year}{2021}), \eprint{2010.10774}.

\bibitem[{\citenamefont{Jia et~al.}(2020)\citenamefont{Jia, Wei, Xiao, and
  Yuan}}]{Jia:2019qbl}
\bibinfo{author}{\bibfnamefont{J.}~\bibnamefont{Jia}},
  \bibinfo{author}{\bibfnamefont{S.-Y.} \bibnamefont{Wei}},
  \bibinfo{author}{\bibfnamefont{B.-W.} \bibnamefont{Xiao}}, \bibnamefont{and}
  \bibinfo{author}{\bibfnamefont{F.}~\bibnamefont{Yuan}},
  \bibinfo{journal}{Phys. Rev. D} \textbf{\bibinfo{volume}{101}},
  \bibinfo{pages}{094008} (\bibinfo{year}{2020}), \eprint{1910.05290}.

\bibitem[{\citenamefont{Jalilian-Marian and
  Rezaeian}(2012{\natexlab{a}})}]{Jalilian-Marian:2012wwi}
\bibinfo{author}{\bibfnamefont{J.}~\bibnamefont{Jalilian-Marian}}
  \bibnamefont{and} \bibinfo{author}{\bibfnamefont{A.~H.}
  \bibnamefont{Rezaeian}}, \bibinfo{journal}{Phys. Rev. D}
  \textbf{\bibinfo{volume}{86}}, \bibinfo{pages}{034016}
  (\bibinfo{year}{2012}{\natexlab{a}}), \eprint{1204.1319}.

\bibitem[{\citenamefont{Jalilian-Marian and
  Rezaeian}(2012{\natexlab{b}})}]{Jalilian-Marian:2011tvq}
\bibinfo{author}{\bibfnamefont{J.}~\bibnamefont{Jalilian-Marian}}
  \bibnamefont{and} \bibinfo{author}{\bibfnamefont{A.~H.}
  \bibnamefont{Rezaeian}}, \bibinfo{journal}{Phys. Rev. D}
  \textbf{\bibinfo{volume}{85}}, \bibinfo{pages}{014017}
  (\bibinfo{year}{2012}{\natexlab{b}}), \eprint{1110.2810}.

\bibitem[{\citenamefont{Jalilian-Marian}(2005)}]{Jalilian-Marian:2005tod}
\bibinfo{author}{\bibfnamefont{J.}~\bibnamefont{Jalilian-Marian}},
  \bibinfo{journal}{Nucl. Phys. A} \textbf{\bibinfo{volume}{753}},
  \bibinfo{pages}{307} (\bibinfo{year}{2005}), \eprint{hep-ph/0501222}.

\bibitem[{\citenamefont{Jalilian-Marian}(2004)}]{Jalilian-Marian:2004cdc}
\bibinfo{author}{\bibfnamefont{J.}~\bibnamefont{Jalilian-Marian}},
  \bibinfo{journal}{Nucl. Phys. A} \textbf{\bibinfo{volume}{739}},
  \bibinfo{pages}{319} (\bibinfo{year}{2004}), \eprint{nucl-th/0402014}.

\bibitem[{\citenamefont{Dumitru et~al.}(2011)\citenamefont{Dumitru,
  Jalilian-Marian, and Petreska}}]{Dumitru:2011zz}
\bibinfo{author}{\bibfnamefont{A.}~\bibnamefont{Dumitru}},
  \bibinfo{author}{\bibfnamefont{J.}~\bibnamefont{Jalilian-Marian}},
  \bibnamefont{and} \bibinfo{author}{\bibfnamefont{E.}~\bibnamefont{Petreska}},
  \bibinfo{journal}{Phys. Rev. D} \textbf{\bibinfo{volume}{84}},
  \bibinfo{pages}{014018} (\bibinfo{year}{2011}), \eprint{1105.4155}.

\bibitem[{\citenamefont{Dumitru and Jalilian-Marian}(2010)}]{Dumitru:2010ak}
\bibinfo{author}{\bibfnamefont{A.}~\bibnamefont{Dumitru}} \bibnamefont{and}
  \bibinfo{author}{\bibfnamefont{J.}~\bibnamefont{Jalilian-Marian}},
  \bibinfo{journal}{Phys. Rev. D} \textbf{\bibinfo{volume}{82}},
  \bibinfo{pages}{074023} (\bibinfo{year}{2010}), \eprint{1008.0480}.

\bibitem[{\citenamefont{Kang et~al.}(2012)\citenamefont{Kang, Vitev, and
  Xing}}]{Kang:2011bp}
\bibinfo{author}{\bibfnamefont{Z.-B.} \bibnamefont{Kang}},
  \bibinfo{author}{\bibfnamefont{I.}~\bibnamefont{Vitev}}, \bibnamefont{and}
  \bibinfo{author}{\bibfnamefont{H.}~\bibnamefont{Xing}},
  \bibinfo{journal}{Phys. Rev. D} \textbf{\bibinfo{volume}{85}},
  \bibinfo{pages}{054024} (\bibinfo{year}{2012}), \eprint{1112.6021}.

\bibitem[{\citenamefont{Kolb\'e et~al.}(2021)\citenamefont{Kolb\'e, Roy,
  Salazar, Schenke, and Venugopalan}}]{Kolbe:2020tlq}
\bibinfo{author}{\bibfnamefont{I.}~\bibnamefont{Kolb\'e}},
  \bibinfo{author}{\bibfnamefont{K.}~\bibnamefont{Roy}},
  \bibinfo{author}{\bibfnamefont{F.}~\bibnamefont{Salazar}},
  \bibinfo{author}{\bibfnamefont{B.}~\bibnamefont{Schenke}}, \bibnamefont{and}
  \bibinfo{author}{\bibfnamefont{R.}~\bibnamefont{Venugopalan}},
  \bibinfo{journal}{JHEP} \textbf{\bibinfo{volume}{01}}, \bibinfo{pages}{052}
  (\bibinfo{year}{2021}), \eprint{2008.04372}.

\bibitem[{\citenamefont{Jalilian-Marian}(2006)}]{Jalilian-Marian:2005qbq}
\bibinfo{author}{\bibfnamefont{J.}~\bibnamefont{Jalilian-Marian}},
  \bibinfo{journal}{Nucl. Phys. A} \textbf{\bibinfo{volume}{770}},
  \bibinfo{pages}{210} (\bibinfo{year}{2006}), \eprint{hep-ph/0509338}.

\bibitem[{\citenamefont{M\"antysaari et~al.}(2020)\citenamefont{M\"antysaari,
  Mueller, Salazar, and Schenke}}]{Mantysaari:2019hkq}
\bibinfo{author}{\bibfnamefont{H.}~\bibnamefont{M\"antysaari}},
  \bibinfo{author}{\bibfnamefont{N.}~\bibnamefont{Mueller}},
  \bibinfo{author}{\bibfnamefont{F.}~\bibnamefont{Salazar}}, \bibnamefont{and}
  \bibinfo{author}{\bibfnamefont{B.}~\bibnamefont{Schenke}},
  \bibinfo{journal}{Phys. Rev. Lett.} \textbf{\bibinfo{volume}{124}},
  \bibinfo{pages}{112301} (\bibinfo{year}{2020}), \eprint{1912.05586}.

\bibitem[{\citenamefont{Boussarie
  et~al.}(2021{\natexlab{b}})\citenamefont{Boussarie, M\"antysaari, Salazar,
  and Schenke}}]{Boussarie:2021ybe}
\bibinfo{author}{\bibfnamefont{R.}~\bibnamefont{Boussarie}},
  \bibinfo{author}{\bibfnamefont{H.}~\bibnamefont{M\"antysaari}},
  \bibinfo{author}{\bibfnamefont{F.}~\bibnamefont{Salazar}}, \bibnamefont{and}
  \bibinfo{author}{\bibfnamefont{B.}~\bibnamefont{Schenke}},
  \bibinfo{journal}{JHEP} \textbf{\bibinfo{volume}{09}}, \bibinfo{pages}{178}
  (\bibinfo{year}{2021}{\natexlab{b}}), \eprint{2106.11301}.

\bibitem[{\citenamefont{Kotko et~al.}(2017)\citenamefont{Kotko, Kutak, Sapeta,
  Stasto, and Strikman}}]{Kotko:2017oxg}
\bibinfo{author}{\bibfnamefont{P.}~\bibnamefont{Kotko}},
  \bibinfo{author}{\bibfnamefont{K.}~\bibnamefont{Kutak}},
  \bibinfo{author}{\bibfnamefont{S.}~\bibnamefont{Sapeta}},
  \bibinfo{author}{\bibfnamefont{A.~M.} \bibnamefont{Stasto}},
  \bibnamefont{and} \bibinfo{author}{\bibfnamefont{M.}~\bibnamefont{Strikman}},
  \bibinfo{journal}{Eur. Phys. J. C} \textbf{\bibinfo{volume}{77}},
  \bibinfo{pages}{353} (\bibinfo{year}{2017}), \eprint{1702.03063}.

\bibitem[{\citenamefont{Salazar and Schenke}(2019)}]{Salazar:2019ncp}
\bibinfo{author}{\bibfnamefont{F.}~\bibnamefont{Salazar}} \bibnamefont{and}
  \bibinfo{author}{\bibfnamefont{B.}~\bibnamefont{Schenke}},
  \bibinfo{journal}{Phys. Rev. D} \textbf{\bibinfo{volume}{100}},
  \bibinfo{pages}{034007} (\bibinfo{year}{2019}), \eprint{1905.03763}.

\bibitem[{\citenamefont{M\"antysaari et~al.}(2019)\citenamefont{M\"antysaari,
  Mueller, and Schenke}}]{Mantysaari:2019csc}
\bibinfo{author}{\bibfnamefont{H.}~\bibnamefont{M\"antysaari}},
  \bibinfo{author}{\bibfnamefont{N.}~\bibnamefont{Mueller}}, \bibnamefont{and}
  \bibinfo{author}{\bibfnamefont{B.}~\bibnamefont{Schenke}},
  \bibinfo{journal}{Phys. Rev. D} \textbf{\bibinfo{volume}{99}},
  \bibinfo{pages}{074004} (\bibinfo{year}{2019}), \eprint{1902.05087}.

\bibitem[{\citenamefont{Altinoluk et~al.}(2016)\citenamefont{Altinoluk,
  Armesto, Beuf, and Rezaeian}}]{Altinoluk:2015dpi}
\bibinfo{author}{\bibfnamefont{T.}~\bibnamefont{Altinoluk}},
  \bibinfo{author}{\bibfnamefont{N.}~\bibnamefont{Armesto}},
  \bibinfo{author}{\bibfnamefont{G.}~\bibnamefont{Beuf}}, \bibnamefont{and}
  \bibinfo{author}{\bibfnamefont{A.~H.} \bibnamefont{Rezaeian}},
  \bibinfo{journal}{Phys. Lett. B} \textbf{\bibinfo{volume}{758}},
  \bibinfo{pages}{373} (\bibinfo{year}{2016}), \eprint{1511.07452}.

\bibitem[{\citenamefont{Dumitru et~al.}(2015)\citenamefont{Dumitru, Lappi, and
  Skokov}}]{Dumitru:2015gaa}
\bibinfo{author}{\bibfnamefont{A.}~\bibnamefont{Dumitru}},
  \bibinfo{author}{\bibfnamefont{T.}~\bibnamefont{Lappi}}, \bibnamefont{and}
  \bibinfo{author}{\bibfnamefont{V.}~\bibnamefont{Skokov}},
  \bibinfo{journal}{Phys. Rev. Lett.} \textbf{\bibinfo{volume}{115}},
  \bibinfo{pages}{252301} (\bibinfo{year}{2015}), \eprint{1508.04438}.

\bibitem[{\citenamefont{Iancu et~al.}(2022)\citenamefont{Iancu, Mueller, and
  Triantafyllopoulos}}]{Iancu:2021rup}
\bibinfo{author}{\bibfnamefont{E.}~\bibnamefont{Iancu}},
  \bibinfo{author}{\bibfnamefont{A.~H.} \bibnamefont{Mueller}},
  \bibnamefont{and} \bibinfo{author}{\bibfnamefont{D.~N.}
  \bibnamefont{Triantafyllopoulos}}, \bibinfo{journal}{Phys. Rev. Lett.}
  \textbf{\bibinfo{volume}{128}}, \bibinfo{pages}{202001}
  (\bibinfo{year}{2022}), \eprint{2112.06353}.

\bibitem[{\citenamefont{Hatta et~al.}(2016)\citenamefont{Hatta, Xiao, and
  Yuan}}]{Hatta:2016dxp}
\bibinfo{author}{\bibfnamefont{Y.}~\bibnamefont{Hatta}},
  \bibinfo{author}{\bibfnamefont{B.-W.} \bibnamefont{Xiao}}, \bibnamefont{and}
  \bibinfo{author}{\bibfnamefont{F.}~\bibnamefont{Yuan}},
  \bibinfo{journal}{Phys. Rev. Lett.} \textbf{\bibinfo{volume}{116}},
  \bibinfo{pages}{202301} (\bibinfo{year}{2016}), \eprint{1601.01585}.

\bibitem[{\citenamefont{Boussarie et~al.}(2016)\citenamefont{Boussarie,
  Grabovsky, Szymanowski, and Wallon}}]{Boussarie:2016ogo}
\bibinfo{author}{\bibfnamefont{R.}~\bibnamefont{Boussarie}},
  \bibinfo{author}{\bibfnamefont{A.~V.} \bibnamefont{Grabovsky}},
  \bibinfo{author}{\bibfnamefont{L.}~\bibnamefont{Szymanowski}},
  \bibnamefont{and} \bibinfo{author}{\bibfnamefont{S.}~\bibnamefont{Wallon}},
  \bibinfo{journal}{JHEP} \textbf{\bibinfo{volume}{11}}, \bibinfo{pages}{149}
  (\bibinfo{year}{2016}), \eprint{1606.00419}.

\bibitem[{\citenamefont{Boussarie et~al.}(2014)\citenamefont{Boussarie,
  Grabovsky, Szymanowski, and Wallon}}]{Boussarie:2014lxa}
\bibinfo{author}{\bibfnamefont{R.}~\bibnamefont{Boussarie}},
  \bibinfo{author}{\bibfnamefont{A.~V.} \bibnamefont{Grabovsky}},
  \bibinfo{author}{\bibfnamefont{L.}~\bibnamefont{Szymanowski}},
  \bibnamefont{and} \bibinfo{author}{\bibfnamefont{S.}~\bibnamefont{Wallon}},
  \bibinfo{journal}{JHEP} \textbf{\bibinfo{volume}{09}}, \bibinfo{pages}{026}
  (\bibinfo{year}{2014}), \eprint{1405.7676}.

\bibitem[{\citenamefont{Braidot}(2011)}]{Braidot:2010ig}
\bibinfo{author}{\bibfnamefont{E.}~\bibnamefont{Braidot}}
  (\bibinfo{collaboration}{STAR}), \bibinfo{journal}{Nucl. Phys. A}
  \textbf{\bibinfo{volume}{854}}, \bibinfo{pages}{168} (\bibinfo{year}{2011}),
  \eprint{1008.3989}.

\bibitem[{\citenamefont{Adare et~al.}(2011)}]{Adare:2011sc}
\bibinfo{author}{\bibfnamefont{A.}~\bibnamefont{Adare}} \bibnamefont{et~al.}
  (\bibinfo{collaboration}{PHENIX}), \bibinfo{journal}{Phys. Rev. Lett.}
  \textbf{\bibinfo{volume}{107}}, \bibinfo{pages}{172301}
  (\bibinfo{year}{2011}), \eprint{1105.5112}.

\bibitem[{\citenamefont{Aschenauer et~al.}(2016)}]{Aschenauer:2016our}
\bibinfo{author}{\bibfnamefont{E.-C.} \bibnamefont{Aschenauer}}
  \bibnamefont{et~al.} (\bibinfo{year}{2016}), \eprint{1602.03922}.

\bibitem[{\citenamefont{Accardi et~al.}(2016)}]{Accardi:2012qut}
\bibinfo{author}{\bibfnamefont{A.}~\bibnamefont{Accardi}} \bibnamefont{et~al.},
  \bibinfo{journal}{Eur. Phys. J. A} \textbf{\bibinfo{volume}{52}},
  \bibinfo{pages}{268} (\bibinfo{year}{2016}), \eprint{1212.1701}.

\bibitem[{\citenamefont{Chirilli
  et~al.}(2012{\natexlab{a}})\citenamefont{Chirilli, Xiao, and
  Yuan}}]{Chirilli:2011km}
\bibinfo{author}{\bibfnamefont{G.~A.} \bibnamefont{Chirilli}},
  \bibinfo{author}{\bibfnamefont{B.-W.} \bibnamefont{Xiao}}, \bibnamefont{and}
  \bibinfo{author}{\bibfnamefont{F.}~\bibnamefont{Yuan}},
  \bibinfo{journal}{Phys. Rev. Lett.} \textbf{\bibinfo{volume}{108}},
  \bibinfo{pages}{122301} (\bibinfo{year}{2012}{\natexlab{a}}),
  \eprint{1112.1061}.

\bibitem[{\citenamefont{Chirilli
  et~al.}(2012{\natexlab{b}})\citenamefont{Chirilli, Xiao, and
  Yuan}}]{Chirilli:2012jd}
\bibinfo{author}{\bibfnamefont{G.~A.} \bibnamefont{Chirilli}},
  \bibinfo{author}{\bibfnamefont{B.-W.} \bibnamefont{Xiao}}, \bibnamefont{and}
  \bibinfo{author}{\bibfnamefont{F.}~\bibnamefont{Yuan}},
  \bibinfo{journal}{Phys. Rev. D} \textbf{\bibinfo{volume}{86}},
  \bibinfo{pages}{054005} (\bibinfo{year}{2012}{\natexlab{b}}),
  \eprint{1203.6139}.

\bibitem[{\citenamefont{Ayala et~al.}(2016)\citenamefont{Ayala, Hentschinski,
  Jalilian-Marian, and Tejeda-Yeomans}}]{Ayala:2016lhd}
\bibinfo{author}{\bibfnamefont{A.}~\bibnamefont{Ayala}},
  \bibinfo{author}{\bibfnamefont{M.}~\bibnamefont{Hentschinski}},
  \bibinfo{author}{\bibfnamefont{J.}~\bibnamefont{Jalilian-Marian}},
  \bibnamefont{and} \bibinfo{author}{\bibfnamefont{M.~E.}
  \bibnamefont{Tejeda-Yeomans}}, \bibinfo{journal}{Phys. Lett. B}
  \textbf{\bibinfo{volume}{761}}, \bibinfo{pages}{229} (\bibinfo{year}{2016}),
  \eprint{1604.08526}.

\bibitem[{\citenamefont{Ayala et~al.}(2017)\citenamefont{Ayala, Hentschinski,
  Jalilian-Marian, and Tejeda-Yeomans}}]{Ayala:2017rmh}
\bibinfo{author}{\bibfnamefont{A.}~\bibnamefont{Ayala}},
  \bibinfo{author}{\bibfnamefont{M.}~\bibnamefont{Hentschinski}},
  \bibinfo{author}{\bibfnamefont{J.}~\bibnamefont{Jalilian-Marian}},
  \bibnamefont{and} \bibinfo{author}{\bibfnamefont{M.~E.}
  \bibnamefont{Tejeda-Yeomans}}, \bibinfo{journal}{Nucl. Phys. B}
  \textbf{\bibinfo{volume}{920}}, \bibinfo{pages}{232} (\bibinfo{year}{2017}),
  \eprint{1701.07143}.

\bibitem[{\citenamefont{Caucal et~al.}(2021)\citenamefont{Caucal, Salazar, and
  Venugopalan}}]{Caucal:2021ent}
\bibinfo{author}{\bibfnamefont{P.}~\bibnamefont{Caucal}},
  \bibinfo{author}{\bibfnamefont{F.}~\bibnamefont{Salazar}}, \bibnamefont{and}
  \bibinfo{author}{\bibfnamefont{R.}~\bibnamefont{Venugopalan}},
  \bibinfo{journal}{JHEP} \textbf{\bibinfo{volume}{11}}, \bibinfo{pages}{222}
  (\bibinfo{year}{2021}), \eprint{2108.06347}.

\bibitem[{\citenamefont{Caucal et~al.}(2022)\citenamefont{Caucal, Salazar,
  Schenke, and Venugopalan}}]{Caucal:2022ulg}
\bibinfo{author}{\bibfnamefont{P.}~\bibnamefont{Caucal}},
  \bibinfo{author}{\bibfnamefont{F.}~\bibnamefont{Salazar}},
  \bibinfo{author}{\bibfnamefont{B.}~\bibnamefont{Schenke}}, \bibnamefont{and}
  \bibinfo{author}{\bibfnamefont{R.}~\bibnamefont{Venugopalan}},
  \bibinfo{journal}{JHEP} \textbf{\bibinfo{volume}{11}}, \bibinfo{pages}{169}
  (\bibinfo{year}{2022}), \eprint{2208.13872}.

\bibitem[{\citenamefont{Bergabo and
  Jalilian-Marian}(2022{\natexlab{a}})}]{Bergabo:2021woe}
\bibinfo{author}{\bibfnamefont{F.}~\bibnamefont{Bergabo}} \bibnamefont{and}
  \bibinfo{author}{\bibfnamefont{J.}~\bibnamefont{Jalilian-Marian}},
  \bibinfo{journal}{Nucl. Phys. A} \textbf{\bibinfo{volume}{1018}},
  \bibinfo{pages}{122358} (\bibinfo{year}{2022}{\natexlab{a}}),
  \eprint{2108.10428}.

\bibitem[{\citenamefont{Bergabo and
  Jalilian-Marian}(2022{\natexlab{b}})}]{Bergabo:2022tcu}
\bibinfo{author}{\bibfnamefont{F.}~\bibnamefont{Bergabo}} \bibnamefont{and}
  \bibinfo{author}{\bibfnamefont{J.}~\bibnamefont{Jalilian-Marian}},
  \bibinfo{journal}{Phys. Rev. D} \textbf{\bibinfo{volume}{106}},
  \bibinfo{pages}{054035} (\bibinfo{year}{2022}{\natexlab{b}}),
  \eprint{2207.03606}.

\bibitem[{\citenamefont{Taels et~al.}(2022)\citenamefont{Taels, Altinoluk,
  Beuf, and Marquet}}]{Taels:2022tza}
\bibinfo{author}{\bibfnamefont{P.}~\bibnamefont{Taels}},
  \bibinfo{author}{\bibfnamefont{T.}~\bibnamefont{Altinoluk}},
  \bibinfo{author}{\bibfnamefont{G.}~\bibnamefont{Beuf}}, \bibnamefont{and}
  \bibinfo{author}{\bibfnamefont{C.}~\bibnamefont{Marquet}}
  (\bibinfo{year}{2022}), \eprint{2204.11650}.

\bibitem[{\citenamefont{Iancu and Mulian}(2021)}]{Iancu:2020mos}
\bibinfo{author}{\bibfnamefont{E.}~\bibnamefont{Iancu}} \bibnamefont{and}
  \bibinfo{author}{\bibfnamefont{Y.}~\bibnamefont{Mulian}},
  \bibinfo{journal}{JHEP} \textbf{\bibinfo{volume}{03}}, \bibinfo{pages}{005}
  (\bibinfo{year}{2021}), \eprint{2009.11930}.

\bibitem[{\citenamefont{Bergabo and
  Jalilian-Marian}(2022{\natexlab{c}})}]{Bergabo:2022zhe}
\bibinfo{author}{\bibfnamefont{F.}~\bibnamefont{Bergabo}} \bibnamefont{and}
  \bibinfo{author}{\bibfnamefont{J.}~\bibnamefont{Jalilian-Marian}}
  (\bibinfo{year}{2022}{\natexlab{c}}), \eprint{2210.03208}.

\bibitem[{\citenamefont{Benic et~al.}(2017)\citenamefont{Benic, Fukushima,
  Garcia-Montero, and Venugopalan}}]{Benic:2016uku}
\bibinfo{author}{\bibfnamefont{S.}~\bibnamefont{Benic}},
  \bibinfo{author}{\bibfnamefont{K.}~\bibnamefont{Fukushima}},
  \bibinfo{author}{\bibfnamefont{O.}~\bibnamefont{Garcia-Montero}},
  \bibnamefont{and}
  \bibinfo{author}{\bibfnamefont{R.}~\bibnamefont{Venugopalan}},
  \bibinfo{journal}{JHEP} \textbf{\bibinfo{volume}{01}}, \bibinfo{pages}{115}
  (\bibinfo{year}{2017}), \eprint{1609.09424}.

\bibitem[{\citenamefont{Dumitru and
  Jalilian-Marian}(2002{\natexlab{a}})}]{Dumitru:2002qt}
\bibinfo{author}{\bibfnamefont{A.}~\bibnamefont{Dumitru}} \bibnamefont{and}
  \bibinfo{author}{\bibfnamefont{J.}~\bibnamefont{Jalilian-Marian}},
  \bibinfo{journal}{Phys. Rev. Lett.} \textbf{\bibinfo{volume}{89}},
  \bibinfo{pages}{022301} (\bibinfo{year}{2002}{\natexlab{a}}),
  \eprint{hep-ph/0204028}.

\bibitem[{\citenamefont{Dumitru and
  Jalilian-Marian}(2002{\natexlab{b}})}]{Dumitru:2001jn}
\bibinfo{author}{\bibfnamefont{A.}~\bibnamefont{Dumitru}} \bibnamefont{and}
  \bibinfo{author}{\bibfnamefont{J.}~\bibnamefont{Jalilian-Marian}},
  \bibinfo{journal}{Phys. Lett. B} \textbf{\bibinfo{volume}{547}},
  \bibinfo{pages}{15} (\bibinfo{year}{2002}{\natexlab{b}}),
  \eprint{hep-ph/0111357}.

\bibitem[{\citenamefont{Ayala et~al.}(1996)\citenamefont{Ayala,
  Jalilian-Marian, McLerran, and Venugopalan}}]{Ayala:1995hx}
\bibinfo{author}{\bibfnamefont{A.}~\bibnamefont{Ayala}},
  \bibinfo{author}{\bibfnamefont{J.}~\bibnamefont{Jalilian-Marian}},
  \bibinfo{author}{\bibfnamefont{L.~D.} \bibnamefont{McLerran}},
  \bibnamefont{and}
  \bibinfo{author}{\bibfnamefont{R.}~\bibnamefont{Venugopalan}},
  \bibinfo{journal}{Phys. Rev. D} \textbf{\bibinfo{volume}{53}},
  \bibinfo{pages}{458} (\bibinfo{year}{1996}), \eprint{hep-ph/9508302}.

\bibitem[{\citenamefont{Balitsky}(1996)}]{Balitsky:1995ub}
\bibinfo{author}{\bibfnamefont{I.}~\bibnamefont{Balitsky}},
  \bibinfo{journal}{Nucl. Phys. B} \textbf{\bibinfo{volume}{463}},
  \bibinfo{pages}{99} (\bibinfo{year}{1996}), \eprint{hep-ph/9509348}.

\bibitem[{\citenamefont{Kovchegov}(2000)}]{Kovchegov:1999yj}
\bibinfo{author}{\bibfnamefont{Y.~V.} \bibnamefont{Kovchegov}},
  \bibinfo{journal}{Phys. Rev. D} \textbf{\bibinfo{volume}{61}},
  \bibinfo{pages}{074018} (\bibinfo{year}{2000}), \eprint{hep-ph/9905214}.

\bibitem[{\citenamefont{Jalilian-Marian
  et~al.}(1997)\citenamefont{Jalilian-Marian, Kovner, Leonidov, and
  Weigert}}]{Jalilian-Marian:1997qno}
\bibinfo{author}{\bibfnamefont{J.}~\bibnamefont{Jalilian-Marian}},
  \bibinfo{author}{\bibfnamefont{A.}~\bibnamefont{Kovner}},
  \bibinfo{author}{\bibfnamefont{A.}~\bibnamefont{Leonidov}}, \bibnamefont{and}
  \bibinfo{author}{\bibfnamefont{H.}~\bibnamefont{Weigert}},
  \bibinfo{journal}{Nucl. Phys. B} \textbf{\bibinfo{volume}{504}},
  \bibinfo{pages}{415} (\bibinfo{year}{1997}), \eprint{hep-ph/9701284}.

\bibitem[{\citenamefont{Jalilian-Marian
  et~al.}(1998{\natexlab{a}})\citenamefont{Jalilian-Marian, Kovner, Leonidov,
  and Weigert}}]{Jalilian-Marian:1997jhx}
\bibinfo{author}{\bibfnamefont{J.}~\bibnamefont{Jalilian-Marian}},
  \bibinfo{author}{\bibfnamefont{A.}~\bibnamefont{Kovner}},
  \bibinfo{author}{\bibfnamefont{A.}~\bibnamefont{Leonidov}}, \bibnamefont{and}
  \bibinfo{author}{\bibfnamefont{H.}~\bibnamefont{Weigert}},
  \bibinfo{journal}{Phys. Rev. D} \textbf{\bibinfo{volume}{59}},
  \bibinfo{pages}{014014} (\bibinfo{year}{1998}{\natexlab{a}}),
  \eprint{hep-ph/9706377}.

\bibitem[{\citenamefont{Jalilian-Marian
  et~al.}(1998{\natexlab{b}})\citenamefont{Jalilian-Marian, Kovner, and
  Weigert}}]{Jalilian-Marian:1997ubg}
\bibinfo{author}{\bibfnamefont{J.}~\bibnamefont{Jalilian-Marian}},
  \bibinfo{author}{\bibfnamefont{A.}~\bibnamefont{Kovner}}, \bibnamefont{and}
  \bibinfo{author}{\bibfnamefont{H.}~\bibnamefont{Weigert}},
  \bibinfo{journal}{Phys. Rev. D} \textbf{\bibinfo{volume}{59}},
  \bibinfo{pages}{014015} (\bibinfo{year}{1998}{\natexlab{b}}),
  \eprint{hep-ph/9709432}.

\bibitem[{\citenamefont{Jalilian-Marian
  et~al.}(1999)\citenamefont{Jalilian-Marian, Kovner, Leonidov, and
  Weigert}}]{Jalilian-Marian:1998tzv}
\bibinfo{author}{\bibfnamefont{J.}~\bibnamefont{Jalilian-Marian}},
  \bibinfo{author}{\bibfnamefont{A.}~\bibnamefont{Kovner}},
  \bibinfo{author}{\bibfnamefont{A.}~\bibnamefont{Leonidov}}, \bibnamefont{and}
  \bibinfo{author}{\bibfnamefont{H.}~\bibnamefont{Weigert}},
  \bibinfo{journal}{Phys. Rev. D} \textbf{\bibinfo{volume}{59}},
  \bibinfo{pages}{034007} (\bibinfo{year}{1999}), \bibinfo{note}{[Erratum:
  Phys.Rev.D 59, 099903 (1999)]}, \eprint{hep-ph/9807462}.

\bibitem[{\citenamefont{Kovner et~al.}(2000)\citenamefont{Kovner, Milhano, and
  Weigert}}]{Kovner:2000pt}
\bibinfo{author}{\bibfnamefont{A.}~\bibnamefont{Kovner}},
  \bibinfo{author}{\bibfnamefont{J.~G.} \bibnamefont{Milhano}},
  \bibnamefont{and} \bibinfo{author}{\bibfnamefont{H.}~\bibnamefont{Weigert}},
  \bibinfo{journal}{Phys. Rev. D} \textbf{\bibinfo{volume}{62}},
  \bibinfo{pages}{114005} (\bibinfo{year}{2000}), \eprint{hep-ph/0004014}.

\bibitem[{\citenamefont{Iancu et~al.}(2001)\citenamefont{Iancu, Leonidov, and
  McLerran}}]{Iancu:2000hn}
\bibinfo{author}{\bibfnamefont{E.}~\bibnamefont{Iancu}},
  \bibinfo{author}{\bibfnamefont{A.}~\bibnamefont{Leonidov}}, \bibnamefont{and}
  \bibinfo{author}{\bibfnamefont{L.~D.} \bibnamefont{McLerran}},
  \bibinfo{journal}{Nucl. Phys. A} \textbf{\bibinfo{volume}{692}},
  \bibinfo{pages}{583} (\bibinfo{year}{2001}), \eprint{hep-ph/0011241}.

\bibitem[{\citenamefont{Ferreiro et~al.}(2002)\citenamefont{Ferreiro, Iancu,
  Leonidov, and McLerran}}]{Ferreiro:2001qy}
\bibinfo{author}{\bibfnamefont{E.}~\bibnamefont{Ferreiro}},
  \bibinfo{author}{\bibfnamefont{E.}~\bibnamefont{Iancu}},
  \bibinfo{author}{\bibfnamefont{A.}~\bibnamefont{Leonidov}}, \bibnamefont{and}
  \bibinfo{author}{\bibfnamefont{L.}~\bibnamefont{McLerran}},
  \bibinfo{journal}{Nucl. Phys. A} \textbf{\bibinfo{volume}{703}},
  \bibinfo{pages}{489} (\bibinfo{year}{2002}), \eprint{hep-ph/0109115}.

\end{thebibliography}
\bibliographystyle{apsrev}

\end{document}